\documentclass[aps,prl,reprint,twocolumn,preprintnumbers,floatfix,nofootinbib]{revtex4-1}

\usepackage[T1]{fontenc} 

\usepackage{graphicx}
\usepackage{epstopdf}
\usepackage{mathrsfs}
\usepackage{amssymb}
\usepackage{verbatim}
\usepackage{color}
\usepackage[dvipsnames]{xcolor}
\usepackage{multirow}
\usepackage{amsmath}
\usepackage{slashed} 
\usepackage{float}
\usepackage[normalem]{ulem}
\usepackage{hyperref}
\usepackage{cancel}
\usepackage{enumerate}
\usepackage{subfig}
\hypersetup{pdfstartview=FitV,colorlinks=true,linkcolor=blue,citecolor=red,filecolor=black,urlcolor=blue,breaklinks=true}

\newcommand{\beq}{\begin{equation}}
\newcommand{\eeq}{\end{equation}}
\newcommand{\bea}{\begin{eqnarray}}
\newcommand{\eea}{\end{eqnarray}}

\usepackage[parfill]{parskip}
\begin{document}\sloppy 

\preprint{UMN--TH--3916/20, FTPI--MINN--20/06}
\preprint{IFT-UAM/CSIC-20-56}

\vspace*{1mm}

\title{Reheating and Post-inflationary Production of Dark Matter}
\author{Marcos A.~G.~Garcia$^{a}$}
\email{marcosa.garcia@uam.es}
\author{Kunio Kaneta$^{b}$}
\email{kkaneta@umn.edu}
\author{Yann Mambrini$^{c}$}
\email{yann.mambrini@ijclab.in2p3.fr}
\author{Keith A. Olive$^{b}$}
\email{olive@umn.edu}
\vspace{0.5cm}

 \affiliation{
$^a$
Instituto de F\'{i}sica Te\'{o}rica (IFT) UAM-CSIC, Campus de Cantoblanco, 28049 Madrid, Spain}

\affiliation{$^b$William I. Fine Theoretical Physics Institute, School of
 Physics and Astronomy, University of Minnesota, Minneapolis, Minnesota 55455,
 USA}
 
\affiliation{${}^c $ Universit\'e Paris-Saclay, CNRS/IN2P3, IJCLab, 91405 Orsay, France}

\date{\today}

\begin{abstract} 
We perform a systematic analysis of dark matter production during post-inflationary reheating.
Following the period of exponential expansion,
the inflaton begins a period of damped oscillations as it decays. These oscillations,
and the evolution of temperature of the thermalized decay products 
depend on the shape of the inflaton potential $V(\Phi)$. We consider potentials of the form,
$\Phi^k$. Standard matter-dominated oscillations occur for $k=2$. In general, the production of dark matter may depend on either (or both) 
the maximum temperature after inflation, or the 
reheating temperature, where the latter is defined when the Universe becomes radiation dominated. 
We show that dark matter production is sensitive to the inflaton potential and depends heavily
on the maximum temperature when $k>2$.
We also consider the production of dark matter with masses larger than the reheating temperature.

\end{abstract}

\maketitle

\setcounter{equation}{0}

\section{I. Introduction}

Since the first computation indicating the presence of a dark component in our Galaxy by Poincar\'e in 1906 \cite{poincare} there were  
 observations of the Coma cluster by Zwicky \cite{Zwicky:1933gu} in 1933 and the analysis of the Andromeda rotation curve by Babcock in 1935 \cite{babcock}, leading to the proposition of a microscopic dark component by Steigman {\it et al.} in 1978 \cite{Gunn:1978gr}. 
However, despite technological developments, and an increase in the size of new generations of experiments on every continent, not a single dark matter (DM) particle has been observed in direct detection experiments \cite{XENON,LUX,PANDAX}.  The WIMP (Weakly Interacting Massive Particle) paradigm appears to be in tension with observations (see \cite{Arcadi:2017kky} for a recent review).
Classic WIMP candidates such 100 GeV neutral particles with standard weak interactions have elastic cross sections which are over 6 orders of magnitude larger than current direct detection limits. 
Indirect detection has been equally unsuccessful.

There are many ``minimal'' extensions of the Standard Model
such as the
Higgs portal \cite{hp1,hp2,hp3,Higgsportal2,Higgsportal3,Higgsportal5} or $Z'$ portals~\cite{Zpportal1,Zpportal2,Zpportal3} that can still evade experimental constraints, but at the price of complexifying the model by introducing new physics above $\simeq 3$ TeV. In this sense, the ``WIMP miracle'' is not as miraculous as it was believed to be in the first place. Even if better motivated, the minimal supersymmetric standard model \cite{Go1983,ehnos} has a large region of its parameter space \cite{cmssm3,cmssm4,mc12,fittino} in tension with LHC results \cite{nosusy1,nosusy2,nosusy3,nosusy4}. 

In this context, it becomes important to look for alternatives. The WIMP miracle is based on the hypothesis of a dark matter particle in thermal equilibrium with the Standard Model over a period of time in the early Universe. 
The dark matter relic density is then independent of initial conditions, and is determined by the freeze-out of annihilations \cite{hlw1,hlw2}. Relaxing this hypothesis opens up interesting cosmological scenarios and potentially new candidates. The popular Feebly Interacting Massive Particle (FIMP) \cite{fimp1,Bernal:2017kxu} paradigm is one of them. The visible and dark sectors can be secluded because of the smallness of their couplings, even Planck-suppressed as in the case of the gravitino \cite{ehnos,gravitino1,gravitino2,gravitino3,gravitino7,gravitino8,gravitino12}. Another possibility is that the two sectors communicate only through the exchange of very massive fields, that may be more massive than the reheating temperature. This is the case in unified SO(10) scenarios \cite{Mambrini:2013iaa,mnoqz1} or anomaly-free U$(1)'$ constructions \cite{Bhattacharyya:2018evo}. It is also possible that both a tiny coupling {\it and} a heavy mediator seclude the visible and dark sectors, as in high-scale supergravity \cite{Benakli:2017whb,grav2,grav3,highsc,Kaneta:2019zgw}, massive spin-2 portal \cite{Bernal:2018qlk} or moduli-portal dark matter \cite{Chowdhury:2018tzw} models.
It is easy to understand that mass-suppressed interactions (either a Planck-suppressed coupling or the exchange of a heavy mediator) generate production rates that are highly dependent on the energy of the primordial plasma. It is  crucial, therefore, to treat the interactions in the early Universe with great care, especially if one wants to take into account non-instantaneous reheating \cite{Giudice:2000ex,Chung:1998rq,Garcia:2017tuj,grav2,Chen:2017kvz} or thermalization \cite{Harigaya:2013vwa,Harigaya:2014waa,Mukaida:2015ria,Garcia:2018wtq,Harigaya:2019tzu} after inflation.

Typically, after the period of exponential expansion has ended, 
the reheating process takes 
place in a matter-dominated background of 
inflaton oscillations. As the inflaton begins to decay, the decay products begin to thermalize and the temperature of this dilute plasma climbs quickly to a maximum temperature, $T_{\rm max}$~\cite{Giudice:2000ex,Chung:1998rq,Garcia:2017tuj,grav2}. Subsequently, the temperature falls as $T \propto a^{-\frac38}$,  where $a$ is the cosmological scale factor, until the Universe becomes dominated by the radiation products at $T_{\rm reh}$.
If the dark matter production cross section scales as $T^n$, the dark matter density is
determined by $T_{\rm reh}$ for $n<6$
and is sensitive to $T_{\rm max}$ for $n\ge 6$. 

In the reheating scenario described above,
it commonly assumed that the inflaton 
undergoes classic harmonic oscillations about a minimum produced by a quadratic potential. If however, the oscillations are anharmonic, and result from a potential other than a quadratic potential, the equation of state during reheating will differ from that 
of a matter-dominated background and will affect the evolution of the thermalization process \cite{Bernal:2019mhf}.

In this paper, we consider, the effect of oscillations produced by a potential of the form  $V(\Phi) = \frac{\lambda}{M^{k-4}}|\Phi|^k$.  These oscillations alter the equation of state during reheating and affect the evolution of temperature as the Universe expands.
It is important to note that for $k\ne2$, the mass of the inflaton is not constant, and hence the change in the equation of state also affects the inflaton decay width, and as a consequence, the evolution of the temperature of the primordial plasma. 
We will show that the resulting dark matter abundance has increased sensitivity to $T_{\rm max}$ when $k>2$. 

It is also possible to produce dark matter with masses in excess of the reheating temperature (so long as its mass is less than $T_{\rm max}$).  As the temperature decreases from $T=T_{\rm max}$, dark matter particles are produced until reheating is complete. However, if the dark matter mass is $m_{\rm DM} > T_{\rm reh}$, production 
ends at $T\simeq m_{\rm DM}$ and the dark matter abundance is suppressed. 

The paper is organized as follows. In Section II we generalize the reheating process in the case of an inflaton potential $V(\Phi) \propto \Phi^k$, analyzing in detail its consequences in non-instantaneous reheating.
In Section III we apply our results to the computation of dark matter production from thermal bath scattering and inflaton decay.
We consider dark matter masses below and above the reheating temperature. 
We present our conclusions in Section IV.

\section{II. The reheating process}

\subsection{The context}

The process of reheating is necessarily model dependent.
It will depend not only on the inflaton potential, but also on the coupling of the inflaton to other fields.  Clearly, some coupling to Standard Model fields is necessary to produce a thermal bath. 
The inflaton may also couple directly to a dark sector,
or dark matter may be produced out of the thermal bath.
Depending on the coupling of the dark matter with the Standard Model, the dark matter may or may not ever come into thermal equilibrium. The reheating process itself may be disassociated from the period of inflation. That is, the part of the potential that 
drives inflation (the exponential expansion) may be distinct from
the part of the potential which leads to a slow reheating process
in which energy stored in scalar field oscillations is 
converted to the thermal bath.

In this paper, we will indeed separate the inflationary era
from reheating. As an example of this type of model, 
we consider T-attractor models~\cite{Kallosh:2013hoa} (described in more detail below).
In these models, the inflationary part of the potential is 
nearly flat as in the Starobinsky model \cite{Staro}. However, there is considerable freedom for the shape of the potential about the minimum. If inflaton decay 
is sufficiently slow, the details of reheating and particle production depend on the potential which controls the oscillatory
behavior of the inflaton and the equation of state during reheating.

We start with the energy density and pressure of a scalar field which can be extracted from the stress-energy tensor, $T_{\mu \nu}$, yielding the standard expressions
\beq
\rho_\Phi = \frac{1}{2} \dot \Phi^2 + V(\Phi);
~~~P_{\Phi} = \frac{1}{2} \dot \Phi^2 - V(\Phi) \, ,
\label{Eq:rhophi}
\eeq
where we have neglected contributions from spatial gradients. 
Conservation of $T_{\mu \nu}$ leads to 
\beq
\dot \rho_\Phi + 3 H(\rho_\Phi + P_\Phi) =0 \, ,
\label{Eq:conservation}
\eeq
\noindent
where $H=\frac{\dot a}{a}$ is the Hubble parameter. Inserting Eq.~(\ref{Eq:rhophi}) into Eq.~(\ref{Eq:conservation}), we obtain the equation of motion for the inflaton
\beq
\ddot \Phi + 3 H \dot \Phi + V'(\Phi)= 0
\label{Eq:motionphi}
\eeq
where $V'(\Phi)=\partial_\Phi V(\Phi)$.

As noted above, we will assume a generic power-law form for the potential about the minimum
\beq\label{eq:Vgen}
V(\Phi) = \lambda \frac{\left|\Phi\right|^k}{M^{k-4}}\, .
\eeq
Here, $M$ is some high energy mass scale, which we can take,
without loss of generality, to be
the Planck scale\footnote{We will use throughout our work $M_P=2.4 \times 10^{18}$ GeV for the reduced Planck mass.} , $M_P$.
This form of the potential can be thought of as the small field limit of T-attractor models~\cite{Kallosh:2013hoa} and
can be derived in no-scale supergravity~\cite{no-scale1,no-scale2,LN}.  The full
potential exhibits Starobinsky-like inflation \cite{Staro} for values of $\Phi > M_P$. More details are given in the Appendix. Note that we use the T-attractor model as a UV-derivable example, but our analysis does not depend at all on the specifics of the example.
The value of $\lambda$ can be fixed 
from the normalization of CMB anisotropies. 
Upon exiting from the inflationary stage, the inflaton
will begin oscillations about the minimum at $\Phi = 0$. \footnote{The absolute value in (\ref{eq:Vgen}) is necessary only for $k=3$ preventing this case from being derived from the supergravity models discussed in the Appendix.}

During the period of inflaton oscillations, the equation of state parameter, $w = P_\Phi/\rho_\Phi$, also oscillates taking values between
$-1$ when $\Phi$ is at its maximum to +1 when $\Phi = 0$. 
It is useful, therefore, to compute an averaged equation of state,
given by $\langle P_\Phi\rangle = w \langle\rho_\Phi\rangle$ 
(see \cite{Ellis:2015pla} for more details). 
Multiplying Eq.(\ref{Eq:motionphi}) by $\Phi$ and taking the mean over one oscillation we obtain 
\beq
\langle \ddot \Phi \Phi \rangle +
\langle \Phi V'(\Phi)\rangle = 0, ~~\Rightarrow 
\langle \dot \Phi^2 \rangle = \langle \Phi V'(\Phi) \rangle\,. 
\label{Eq:dotphi}
\eeq
One deduces from Eq.(\ref{Eq:rhophi})
\begin{eqnarray}
\langle\rho_\Phi\rangle & \;=\; & \left(\frac{k}{2}+1\right) \lambda \frac{\langle\Phi^k\rangle}{M_P^{k-4}}\,, \nonumber \\
\langle P_\Phi\rangle & \;=\; & \left(\frac{k}{2}-1\right) \lambda \frac{\langle\Phi^k\rangle}{M_P^{k-4}}\, ,
\end{eqnarray}
so that
\beq\label{eq:wk}
w = \frac{\langle P_\Phi\rangle}{\langle\rho_\phi\rangle} = \frac{k-2}{k+2}.
\eeq

If we allow the possibility for the inflaton to decay with a width $\Gamma_\Phi$, we can then rewrite Eq.(\ref{Eq:conservation}) as
\beq
\dot \rho_\Phi 
+ 3 \left( \frac{2k}{k+2}\right)
H \rho_\Phi = - \Gamma_\Phi \rho_\Phi.
\label{Eq:eqrhophi}
\eeq
Note that while we use the average equation of state (\ref{eq:wk})
in the evolution of the energy density, it is sufficient (and simpler) to use
the energy density given entirely from the potential. That is using the amplitude or envelope of the oscillations. 

Before looking at the detailed production of dark matter in a universe dominated by the density of energy $\rho_\Phi$, it will be useful to discuss the process of reheating 
in the case of a generic potential $V(\Phi) = \lambda |\Phi|^k/M_P^{k-4}$.

\subsection{The process of reheating}

After inflation ends, the inflaton undergoes a damped (anharmonic) oscillation about its minimum, due to Hubble friction and its decay into light particles (radiation). The evolution of the energy density of this radiation, $\rho_R$, and thus of the instantaneous temperature\footnote{Throughout this paper we assume that the decay products of the inflaton thermalize instantaneously after they are produced.} $T$, as a function of time (or the scale factor $a$) is determined by the solution of the following set of Boltzmann-Friedmann equations
\bea
&&
\dot \rho_R + 4 H \rho_R \;=\; \Gamma_\Phi \rho_\Phi\,,
\label{Eq:eqrhor}
\\
&&
H^2 \;=\; \frac{\rho_\Phi + \rho_R}{3 M_P^2} \;\simeq\; \frac{\rho_\Phi}{3 M_P^2}\, ,
\label{Eq:eqh}
\eea
%
in addition to Eq.~(\ref{Eq:eqrhophi}).
The approximate equality in (\ref{Eq:eqh}) applies to a universe dominated by the inflaton field ($\rho_\Phi \gg \rho_R$), as is true in the early stages of reheating. Although we will make use of this approximation in our analytical computations, we do not impose it in our numerical analysis. The key aspect of our treatment of reheating consists in the realization that, for $k\neq 2$, the inflaton decay rate is not constant in time.

Assuming an effective coupling of the inflaton to Standard Model fermions $f$ of the form $y \Phi \bar f f$, we can write\footnote{A more careful analysis reveals that the decay rate of $\Phi$, obtained by averaging over one oscillation the damping rate of the energy density of the oscillating inflaton condensate, corrects this expression by an $\mathcal{O}(1)$ factor, weakly dependent on $k$~\cite{Shtanov:1994ce,Ichikawa:2008ne}. We omit it from our analysis for simplicity, though it is included in our numerical results. Our main conclusions are unaffected by this omission.} 
\beq
\Gamma_\Phi \;=\; \frac{y^2}{8 \pi} m_\Phi(t)\,,
\label{Eq:eqgammaphi}
\eeq
where the {\em effective mass} $m_\Phi(t)$ is a function of time (and thus of the temperature of the thermal bath). In the adiabatic approximation\footnote{We will not consider the violations of adiabaticity that occur at the time scale of the oscillation of the inflaton, which is much shorter than the duration of reheating.} it can be written as
\begin{align}
m_\Phi^2 \;&\equiv\; \partial^2_\Phi V(\Phi) \nonumber \\
&=\; \lambda ~k(k-1) \Phi^{k-2} M_P^{4-k}
\label{Eq:mphi} \nonumber \\
&=\; k(k-1)M_P^{\frac{2(4-k)}{k}}\lambda^{\frac{2}{k}}\rho_\Phi^{\frac{k-2}{k}}\,.
\end{align}
To arrive at the expression for the effective mass in terms of $\rho_{\Phi}$ we are using an ``envelope'' approximation for $\rho_\Phi$. This approximation is defined in the following way: one may approximate the oscillating inflaton as $\Phi(t) \simeq \Phi_0(t)\cdot \mathcal{P}(t)$. The function $\mathcal{P}$ is periodic and encodes the (an)harmonicity of the short time-scale oscillations in the potential, while the envelope $\Phi_0(t)$ encodes the effect of redshift and decay, and varies on longer time scales. The {\em instantaneous} value of $\Phi_0$ satisfies the equation~\cite{Shtanov:1994ce}
\beq\label{eq:envelope}
V(\Phi_0(t)) \;=\; \rho_{\Phi}(t)\,.
\eeq
Using the envelope is advantageous because one can then immediately ignore short time scales in the analysis, in particular for the effective mass.

In order to study any (particle production) process during reheating, it is indispensable to know the value of the temperature of the radiation bath at any moment of time (or scale factor $a$). At early times, when the Universe is dominated by inflaton oscillations, we find the solution $\rho_\Phi= \rho_\Phi(a)$ from Eq.~(\ref{Eq:eqrhophi}) and subsequently implement it in Eq.~(\ref{Eq:eqrhor}) to determine the evolution $\rho_R=\rho_R(a)$ and therefore $T=T(a)$.

In the early stages of the
reheating, the decay rate of the inflaton is much smaller than the expansion rate $H$ ($\Gamma_\Phi \ll H$). The right-hand side of Eq.~(\ref{Eq:eqrhophi}) can then be neglected, and straightforward integration then gives
\beq
\rho_\Phi(a) \;=\; \rho_{\rm end} \left( \frac{a}{a_{\rm end}} \right)^{-\frac{6k}{k+2}}\,,
\label{Eq:solrhophi}
\eeq
%
where $\rho_{\rm end}$ and $a_{\rm end}$ denote the energy density and scale factor at the end of inflation, respectively.
While the latter ($a_{\rm end}$) is simply a reference point for the scale factor, the value of $\rho_{\rm end}$
does enter into our physical results.
It is defined as the energy density at the moment when the slow roll parameter, $\epsilon = 2 M_P^2 (H'(\Phi)^2/H(\Phi)^2) = 1$ or when $w = -1/3$ \cite{Ellis:2015pla}. At that moment, $\rho_{\rm end} = \frac32 V(\Phi_{\rm end})$, and clearly depends on the potential. In the Appendix, we compute $\Phi_{\rm end}$
for the T-attractor model~\cite{Kallosh:2013hoa} as a function of $k$.

For $k=2$, we recover the classical evolution of a dust-dominated universe
($\rho_\Phi \propto a^{-3}$), whereas for $k=4$ we are in the presence of a ``radiation-like inflaton''-dominated universe ($\rho_\Phi \propto a^{-4}$). This difference in behavior will have dramatic consequences on the temperature evolution and the production of dark matter. 
Substitution of the decay rate (\ref{Eq:eqgammaphi}) and the effective mass (\ref{Eq:mphi}) into (\ref{Eq:eqrhor}), together with the solution for $\rho_{\Phi}$ (\ref{Eq:solrhophi}) we obtain
\begin{align}
\rho_R(a) \;=\; &\frac{y^2}{8 \pi}\sqrt{3k(k-1)}\lambda^{\frac{1}{k}}
M_P^{\frac{4}{k}} 
\left(\frac{k+2}{14-2k} \right)
\nonumber
\\
&\times \rho_{\rm end}^{\frac{k-1}{k}}
\left( \frac{a_{\rm end}}{a} \right)^4 
\left[
\left(\frac{a}{a_{\rm end}} \right)^{\frac{14-2k}{k+2}}-1
\right]\,.
\label{Eq:rhorfinal}
\end{align}
Note that the dependence of $\rho_R$ on $a$ found here is very different from that in \cite{Bernal:2019mhf}, where 
$\rho_R \sim a^{-3k/(k+2)}$ compared with $\rho_R \sim a^{(6-6k)/(k+2)}$ in Eq.~(\ref{Eq:rhorfinal}), though the two expressions agree for $k=2$. This is presumably because the decay width was held fixed in \cite{Bernal:2019mhf}, whereas for $k>2$, any width proportional to the inflaton mass will vary with its evolution. 

In thermal equilibrium, the temperature of the inflaton decay products will be simply given by
\beq \label{Eq:temperature}
T(a) \;=\; \left(\frac{30 \rho_R(a)}{\pi^2 g_*}\right)^{\frac{1}{4}}\,,
\eeq
where $g_*$ denotes the effective number of relativistic degrees of freedom.
Note that for $a\gg a_{\rm end}$, 
\beq
T \;\propto\; a^{-\frac{3k-3}{2k+4}}\,.
\label{scale}
\eeq
For $k=2$, we recover the well-known $T\propto a^{-\frac{3}{8}}$ for the redshift of the temperature during dust-like reheating~\cite{Giudice:2000ex,Chung:1998rq,Garcia:2017tuj}. 

\begin{figure}[t!]
\includegraphics[width=1.0\columnwidth]{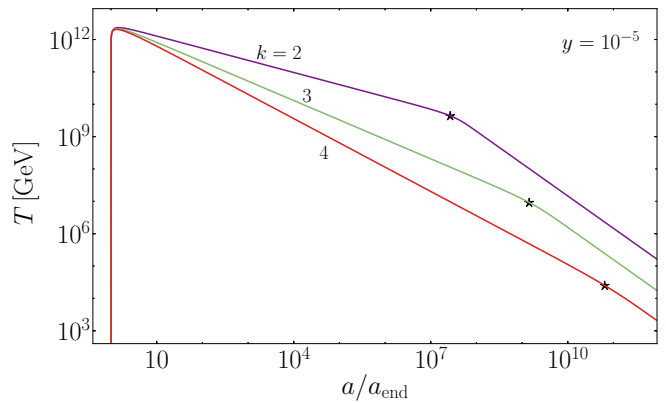}
\caption{Scale-factor dependence of the instantaneous temperature during reheating for selected values of $k$ with $y=10^{-5}$. The scale factor is rescaled by its value at the end of inflation. The star signals inflaton-radiation equality, $\rho_{\Phi}=\rho_R$, corresponding to $T=T_{\rm reh}$.
}
\label{fig:TgPlotA}
\end{figure}

Note that for larger $k$, the temperature has a steeper dependence on the scale factor, e.g.~$T\propto a^{-\frac{3}{4}}$ for radiation-like reheating with $k=4$. 
Indeed, in this case the energy density of $\Phi$, $\rho_\Phi$, redshifts as $a^{-4}$ (\ref{Eq:solrhophi}), faster than for a dust-like inflaton where $\rho_\Phi \propto a^{-3}$. 
This is to be expected as for $k=4$, $\Phi$ is massless at the minimum and evolves as radiation.
Subsequently, this radiation will be further redshifted by expansion. The temperature in the bath is, in a sense, doubly redshifted (production + expansion) compared to a dust-like inflaton decay.
Figure \ref{fig:TgPlotA} exemplifies this steeper redshift of the temperature during reheating for $k=3$ and $4$, compared to $k=2$. 
As a consequence, for $k=4$ the Universe begins to be dominated by the radiation when its scale factor is 3 orders of magnitude larger than it would be for a dust-like inflaton ($k=2$). The effect is anything but an anecdote, as it corresponds to more than 5 orders of magnitude of difference in $T_{\rm reh}$. This comes from the fact that the decay width of the inflaton, proportional to $m_\Phi(t)$, decreases with time for $k=4$. We will explain this phenomena in detail in a dedicated section, below. We note
that for $k=4$ and particularly for low $y$,
the reheating temperature may drop so low as to be 
problematic with baryogenesis, and perhaps nucleosynthesis. We comment further on this possibility in the Appendix.

\subsection{The maximum and reheating temperatures}

Inflation leads to a cold, empty universe that is re-populated during reheating. In the instantaneous thermalization approximation, the temperature of the radiation plasma initially grows until it reaches a maximum temperature $T_{\rm max}$, after which it decreases to the temperature at the end of reheating, $T_{\rm reh}$, and below (see Fig.~\ref{fig:TgPlotA}).

In order to calculate the maximum temperature, one must first compute the scale factor $a_{\rm max}$ for which the temperature (\ref{Eq:temperature}) or the energy density (\ref{Eq:rhorfinal}) is maximized. From this procedure, we obtain that
\beq
a_{\rm max}=a_{\rm end}\left( \frac{2 k + 4}{3k-3} \right)^{\frac{k+2}{14-2k}}\,.
\eeq
This in turn implies that
\begin{align} \notag
T_{\rm max}^4 \;=\; &\frac{15 y^2}{16 g_* \pi^3} \sqrt{3 k(k-1)}\lambda^{\frac{1}{k}} M_P^{\frac{4}{k}}   \\
& \times \rho_{\rm end}^{\frac{k-1}{k}}  \left(\frac{3k -3}{2k+4} \right)^{\frac{3(k-1)}{7-k}}\,.
\label{Eq:tmax}
\end{align}
To compute $T_{\rm max}$, we must specify the potential
to determine $\rho_{\rm end}$ which is discussed in more detail for the T-attractor model~\cite{Kallosh:2013hoa} in the Appendix.
The value of $\lambda$ is set from the normalization of CMB anisotropies and also depends on $k$ as further discussed in the Appendix. For the case of the T-attractor model,
the monomial in Eq. (\ref{eq:Vgen}) is a good approximation
and numerically we find very similar values for $T_{\rm max}$ for all three cases, $k = 2,3$ and 4 (particularly when plotted on a log scale as in Fig.~\ref{fig:TgPlotA}).

To compute the reheating temperature, one must first define what signals the end of reheating. We consider that reheating ends when the radiation density begins to dominate over the inflaton density, i.e.~when $\rho_\Phi = \rho_R$.\footnote{Note that under this definition, the production of entropy from inflaton decay continues for some time after the end of reheating.} From Eqs.~(\ref{Eq:solrhophi}) and (\ref{Eq:rhorfinal}) we find the following approximation for the scale factor at equality,
\begin{align}\notag 
\left(\frac{a_{\rm reh}}{a_{\rm end}}\right)^{\frac{3k}{k+2}} \;\simeq\; &\left(\frac{16\pi (7-k)}{(k+2)\sqrt{3k(k-1)}}\right)^{k/2}\\ \label{eq:arehAn}
&\times \left(\frac{\rho_{\rm end}}{\lambda M_P^4}\right)^{1/2}  y^{-k}\,,
\end{align}
from which we determine the reheating temperature,
\beq\label{eq:TrehAn}
T_{\rm reh}^4 \;=\; \frac{15\left[3k(k-1)\right]^{\frac{k}{2}}  y^{2k} \lambda }{2^{4k-1}  \pi^{2+k} g_* }
 \left(\frac{k+2}{7-k} \right)^k
M_P^4 \,.
\eeq
As one can see, the dependence on $\rho_{\rm end}$ has disappeared from $T_{\rm reh}$. 
For the T-attractor model described in the Appendix,
for $k = (2,3,4)$ and $y = 10^{-5}$, we find $\lambda = (2,0.9,0.3)\times 10^{-11}$ and $T_{\rm reh} = (4.1 \times 10^9,1.0 \times 10^7, 3.2 \times 10^4)$ GeV,
respectively. 

Note that this definition of the reheating temperature
is slightly different from the more common definition
used in instantaneous reheating where the moment of reheating 
is defined to be $\Gamma = \frac32 H$, which is sensible for exponential decay but loses meaning for $k > 2$. 
The ratio of $T_{\rm reh}^4$ from $\Gamma = \frac32 H$ to the expression in Eq.~(\ref{eq:TrehAn}) is
$(2^{5k-6}/3^{2k-2}) ((7-k)/(k+2))^k$. 
For $k= 2$, this means that $T_{\rm reh}$
from Eq.~(\ref{eq:TrehAn}) is smaller by a factor of $\sqrt{3/5}$
relative to that used in instantaneous reheating.
The definition used in this paper is better suited for generic $k$.

We show in Fig.~\ref{fig:TgPlotA} the points (marked by stars) where the Universe begins to be dominated by radiation ($\rho_\Phi = \rho_R$). 
Note that the steeper scale-factor dependence leads to a lower $T_{\rm reh}$ for larger $k$ for a given inflaton-matter coupling. To be more precise, this comes from the fact that reheating is
{\it delayed} for larger values of $k$, delay which implies lower values of $\rho_\Phi$ (and as a consequence $\rho_R$) at reheating. The decay width, given in Eq.~(\ref{Eq:eqgammaphi})
is proportional to $m_\Phi(t)$. While $m_\Phi(t)$ is constant
for $k=2$, it decreases with time for $k>2$. The smaller decay rate causes the delay in reheating and thus results in a lower 
temperature, $T_{\rm reh}$. 
On the other hand, the maximum temperature is only weakly dependent on $k$ and smaller only by a factor of $\sim 0.9$ for $k=4$ relative to the value at $k=2$, $T_{\rm max}\simeq 2.3\times 10^{12}\,{\rm GeV}$. This is because the $k$ dependence in $\rho_{\rm end}$ nearly cancels the explicit $k$ dependence in Eq.~(\ref{Eq:tmax}) and the implicit dependence in $\lambda$.

\begin{figure}[t]
{\includegraphics[width=1.0\columnwidth]{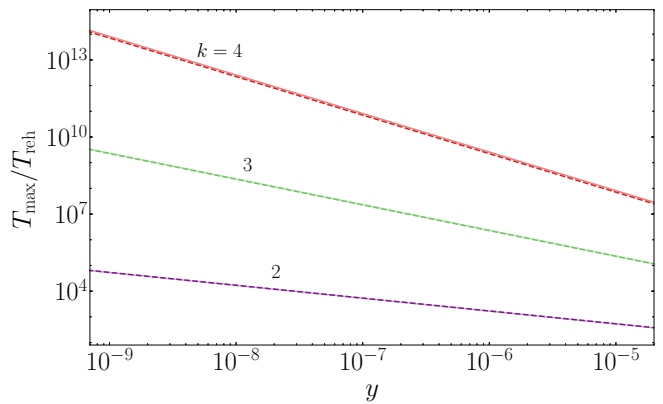}}
\caption{The ratio of the maximum and reheating temperatures as a function of the Yukawa coupling $y$, for selected values of $k$. Continuous: fully numerical solution for the T-attractor model (\ref{eq:Vattractor}). Dashed: the approximations (\ref{eq:Tratk2})-(\ref{eq:Tratk4}).}
\label{fig:TratioY}
\end{figure}

In the following section, which contains our computation of the dark matter abundance, the ratio $T_{\rm max}/T_{\rm reh}$ will play a key role. This ratio is shown in Fig.~\ref{fig:TratioY} as a function of the Yukawa coupling $y$ for $k=2,3,4$. 
We show both our analytic solution for the ratio given in Eqs.~(\ref{eq:Tratk2})-(\ref{eq:Tratk4}) below (dashed) and full
numerical result (solid). 
As one can see, the numerical analysis\footnote{Throughout our work, the numerical results are obtained by solving the full Boltzmann-Friedmann system (\ref{Eq:eqrhophi})-(\ref{Eq:eqh}).} is in perfect agreement with the analytical solutions, i.e., 
$T_{\rm max}/T_{\rm reh}\propto y^{\frac{1-k}{2}}$.

For the selected values of $k$, the following simple expressions can be derived:
\begin{align}\label{eq:Tratk2}
\frac{T_{\rm max}}{T_{\rm reh}}\Bigg|_{k=2} \;&=\; 
\sqrt{\frac{5}{y}}
\left( \frac{3}{2^{33}} \right)^{\frac{1}{40}}
\left(\frac{\pi^2 \rho_{\rm end}}{\lambda M_P^4} \right)^{\frac{1}{8}} \,,
\\ \label{eq:Tratk3}
\frac{T_{\rm max}}{T_{\rm reh}}\Bigg|_{k=3} \;&=\; 
\frac{8}{y}\left( \frac{1}{3 \times 5^9} \right)^{\frac{1}{8}} 
\left( \frac{ \pi^{3} \rho_{\rm end}}{\lambda M_P^4} \right)^{\frac{1}{6}} \,,
\\ \label{eq:Tratk4}
\frac{T_{\rm max}}{T_{\rm reh}}\Bigg|_{k=4} \;&=\; 
\frac{1}{\sqrt{2 y^3}} 
\left( \frac{ \pi^4 \rho_{\rm end}}{\lambda M_P^4} \right)^{\frac{3}{16}} \,.
\end{align}
As noted earlier, the value of $\rho_{\rm end} = \frac32 V(\Phi_{\rm end})$ depends on the potential and the type of inflationary model we consider. As an example, in Fig.~\ref{fig:TratioY},
we use the value of $\Phi_{\rm end}$ given by Eq.~(\ref{Eq:phiend}), corresponding to the T-attractor \cite{Kallosh:2013hoa} completion of the potential in Eq.~(\ref{eq:Vgen}) at large field values (see the Appendix for more details where values of $\rho_{\rm end}/\lambda$ are given).

As one can see, the ratio $T_{\rm max}/T_{\rm reh}$
depends strongly on the value of $k$.
For $y=10^{-5}$, 
we find $T_{\rm max}/T_{\rm reh} = (540, 2.3 \times 10^5, 7.1 \times 10^7)$ for $k = (2, 3, 4)$ using the values of $\rho_{\rm end}/\lambda$ given in the Appendix. 
The ratio is larger by five orders of magnitude for $k=4$ than for $k=2$. We emphasize that this is not an enhancement in $T_{\rm max}$, but rather a reduction in the value of $T_{\rm reh}$ for $k>2$ for a given value of $y$, as we discussed previously when commenting on Fig.~\ref{fig:TgPlotA}. 

We have avoided extrapolating our results for the temperature ratio in Fig.~\ref{fig:TratioY} for $y\gtrsim 10^{-5}$. For any value of the inflaton-SM coupling, the decay products $f$ acquire time-dependent masses induced by the oscillating inflaton background, $m_f=y\,\Phi$. For $y\gtrsim 10^{-5}$ and/or $k>4$, one finds in general that $m_f^2/m_{\Phi}^2 \gtrsim 1$ at some stage of reheating. The perturbative decay of the inflaton can therefore become kinematically suppressed, or dominated by non-perturbative particle production. We leave the detailed study of DM production and reheating beyond these bounds for future work.

\subsection{The Hubble parameter}

We conclude this section by finding an explicit expression for the Hubble parameter as a function of the temperature during reheating. This relation will aid our computation of the DM relic abundance in the following section. Substitution of (\ref{Eq:solrhophi}) and (\ref{Eq:temperature}) into the Friedmann equation (\ref{Eq:eqh}) gives
\beq
H=\frac{\sqrt{\rho_\Phi(T_{\rm reh})}}{\sqrt{3}M_P} \left( \frac{T}{T_{\rm reh}}\right)^{\frac{2k}{k-1}}
\label{Eq:hubble}
\eeq
for $a_{\rm reh}> a \gg a_{\rm end}$. Note that for $k=2$ the previous expression recovers the well-known result $H \propto T^4$, whereas one obtains $H \propto T^{\frac{8}{3}}$ in the case $k=4$. This observation further confirms that for radiation-like reheating, the temperature decreases faster than in the dust-like scenario: for a given value of $H$ we have a larger temperature for larger $k$. 

As a function of time, the Hubble parameter takes the simple form
\beq\label{eq:hubblet}
H \;\simeq\; \frac{k+2}{3kt}\,,
\eeq
for $a_{\rm end}<a<a_{\rm reh}$, which is the classical result for a universe dominated by a homogeneous fluid with equation of state (\ref{eq:wk}).

\section{III. Dark matter production}

In the very early Universe, dark matter can be produced by the scattering 
 of Standard Model particles or by the decay of the inflaton. We review both
possibilities below. Because there is a period of time during which the temperature of the thermal bath exceeds $T_{\rm reh}$, it is possible to produce dark matter particles with mass $m_{\rm DM} > T_{\rm reh}$, and we
consider this possibility as well.

\subsection{DM from thermal bath scattering}

The DM number density, which we will simply denote by $n$, corresponds to the solution of the Boltzmann equation
\beq
\frac{dn}{dt} + 3Hn \;=\; R(t)\,,
\eeq
where $R(t)$ denotes the production rate of DM (per unit volume per unit time). This rate contains the contribution from scatterings in the plasma as well the contribution from the direct decay of the inflaton into DM.
Depending on the magnitude of $R$ compared to $H n$,
dark matter may or may not ever come into thermal equilibrium.  For small $R$, DM remains out of equilibrium as in the case of gravitino production in supersymmetric models \cite{ehnos,gravitino2,gravitino3} and in many generic freeze-in models \cite{fimp1}. 
Making use of (\ref{Eq:hubble}) and (\ref{eq:hubblet}), we can rewrite the Boltzmann equation in terms of the instantaneous temperature as follows,
\beq \label{eq:DMBoltzmann}
\frac{dn}{dT} + \left(\frac{2k+4}{1-k}\right) \frac{n}{T} \;=\; \left(\frac{2k+4}{3-3k}\right) \frac{R(T)}{T H(T)}\,.
\eeq
Equivalently, if we introduce the DM yield $Y\equiv n T^{-\frac{2k+4}{k-1}}$, we can write 
\beq
\frac{dY}{dT} \;=\; - \frac{\sqrt{10}}{\pi \sqrt{g_*}} \left(\frac{2k+4}{k-1} \right) M_P T_{\rm reh}^{\frac{2}{k-1}} T^{\frac{5k+3}{1-k}}~R(T).
\label{eq:DMBoltzmannbis}
\eeq

We parametrize the production rate from out-of-equilibrium scatterings in the following way\footnote{Note that this parametrization corresponds to a thermally a\-ve\-ra\-ged effective cross section $\langle \sigma v\rangle\propto T^n/\Lambda^{n+2}$.}:
\beq\label{eq:RTs}
R^s(T) \;=\; \frac{T^{n+6}}{\Lambda^{n+2}}\,.
\eeq
Here the superscript $s$ denotes production via scatterings in the plasma, and the mass scale $\Lambda$ is identified with the beyond the Standard Model scale of the microscopic model under consideration. Note that this effective description is valid for the duration of reheating provided that $\Lambda\gtrsim T_{\rm max}$. The suppression by the UV scale ensures that DM annihilation can be neglected. Integration of (\ref{eq:DMBoltzmannbis}) after substitution of (\ref{eq:RTs}) leads to the following results:
\begin{itemize}
\item For $n<\frac{10-2k}{k-1}$,
\beq
n^s(T_{\rm reh}) \;=\; \sqrt{\frac{10}{g_*}} \frac{M_P}{\pi}
\frac{2k+4}{n-nk+10-2k} \frac{T_{\rm reh}^{n+4}}{\Lambda^{n+2}}\,.
\label{nlt}
\eeq
\item For $n=\frac{10-2k}{k-1}$,
\beq
n^s(T_{\rm reh}) \;=\; \sqrt{\frac{10}{g_*}}\frac{M_P}{\pi}
\left( \frac{2k+4}{k-1} \right) \frac{T_{\rm reh}^{n+4}}{\Lambda^{n+2}} 
\ln \left(\frac{T_{\rm max}}{T_{\rm reh}} \right)\,.
\label{neq}
\eeq
\item For $n>\frac{10-2k}{k-1}$,
\begin{align}\notag 
n^s(T_{\rm reh}) \;=\; \sqrt{\frac{10}{g_*}}& \frac{M_P}{\pi}
\frac{2k+4}{kn-n-10+2k} \\
&\times
\left( \frac{T_{\rm reh}}{T_{\rm max}}\right)^{\frac{2k+6}{k-1}}
\frac{T_{\rm max}^{n+4}}{\Lambda^{n+2}}\,.
\label{ngt}
\end{align}
\end{itemize}

Note that these results are a generalization 
of \cite{Garcia:2017tuj,Kaneta:2019zgw} applicable to the monomial potential given in Eq.~(\ref{eq:Vgen}) after inflation. For the typical potential with $k=2$, 
i.e.~oscillations of a massive inflaton, the density of dark matter is mainly sensitive to the reheating temperature
if $n < 6$, whereas it is mainly sensitive to the maximum temperature prior to the end of reheating if $n > 6$.
We see that for $k=4$, dark matter production is sensitive to $T_{\rm max}$ for $n \geq 1$. This means that we expect significant production of dark matter in many models. For example, in models where the dark and visible sectors are connected by massive mediators as in SO(10) \cite{fimp3,Mambrini:2013iaa,Nagata:2015dma,Mambrini:2016dca,fimp3,mnoqz1,mnoqz2} or moduli-portal models \cite{Chowdhury:2018tzw}, the production and final density of dark matter will be sensitive to the post-inflationary scalar potential.

The DM number density produced by scatterings in the plasma
given in Eqs.~(\ref{nlt})-(\ref{ngt}) can be converted to the DM contribution to the critical density using
\begin{align}
\Omega^s_{\rm DM}h^2 \;&=\; \frac{m_{\rm DM} n^s(T_0)}{\rho_c h^{-2}} \nonumber \\
&=\; \frac{\pi^2 g_{*s}(T_0) m_{\rm DM} n_{\gamma}(T_0) n^s(T_{\rm reh})}{2\zeta(3)g_{*s}(T_{\rm reh}) T_{\rm reh}^3 \rho_c h^{-2}} \nonumber \\
&=\; 5.9 \times 10^6 {\rm GeV}^{-1} \frac{m_{\rm DM} n^s(T_{\rm reh})}{T_{\rm reh}^3}
\label{oh2s}
\,,
\end{align}
where $g_{*s}(T_0)=43/11$ is the present number of effective relativistic degrees of freedom for the entropy density, $n_{\gamma}(T_0)\simeq 410.66\,{\rm cm}^{-3}$ is the number density of CMB photons, and $\rho_c h^{-2}\simeq 1.0534\times 10^{-5}{\rm GeV\,cm}^{-3}$ is the critical density of the Universe~\cite{Tanabashi:2018oca}. We take $g_{*s}(T_{\rm reh})=g_*(T_{\rm reh})=g_{\rm reh}$, and consider for definiteness the high-temperature Standard Model value $g_{\rm reh}=427/4$. 

\subsection{Production from scattering when $m_{\rm DM} > T_{\rm reh}$}

In the above derivation of $\Omega^s_{\rm DM}$, we have implicitly assumed that $m_{\rm DM} < T_{\rm reh}$, so that the limits of integration of the Boltzmann equation (\ref{eq:DMBoltzmannbis}) ranged from $T_{\rm max}$ to $T_{\rm reh}$. 
For, $m_{\rm DM} > T_{\rm reh}$,
we must cut off the integral at $m_{\rm DM}$.  However, at $T = m_{\rm DM}$, $\rho_R < \rho_\Phi$, and the density of DM matter
will be further diluted by the subsequent decays of the inflaton. Therefore, we compute $n^s(m_{\rm DM}$) and scale it to
$T_{\rm reh}$ using Eq.~(\ref{scale}).
For $n \le (10-2k)/(k-1)$, we find the following:
\begin{itemize}
\item For $n<\frac{10-2k}{k-1}$,
\begin{align}\notag 
n^s(T_{\rm reh}) \;=\;
\sqrt{\frac{10}{g_*}}& \frac{M_P}{\pi}
\frac{2k+4}{n-nk+10-2k} \\
&\times
\left(\frac{T_{\rm reh}}{m_{\rm DM}} \right)^{\frac{2k+6}{k-1}} \frac{m_{\rm DM}^{n+4}}{\Lambda^{n+2}}\,.
\label{nlt2}
\end{align}
\item For $n=\frac{10-2k}{k-1}$,
\begin{align}\notag 
n^s(T_{\rm reh}) \;=\; \sqrt{\frac{10}{g_*}}& \frac{M_P}{\pi}
\left( \frac{2k+4}{k-1} \right) \\ 
&\times 
\left(\frac{T_{\rm reh}}{m_{\rm DM}} \right)^{n+4} \frac{m_{\rm DM}^{n+4}}{\Lambda^{n+2}} 
\ln \left(\frac{T_{\rm max}}{m_{\rm DM}} \right)\,.
\label{neq2}
\end{align}
\end{itemize}
Note that for $n > (10-2k)/(k-1)$,
the result in Eq.~(\ref{ngt}) is unchanged.\footnote{When $T_{\rm reh}<m_{\rm DM}$, the condition $T_{\rm reh}<T_f$ may also be satisfied, where $T_f$ denotes the freeze-out temperature for a thermal (WIMP-like) dark matter candidate. In this case, the abundance of dark matter from freeze-out is reduced~\cite{Roszkowski:2014lga}.}

\subsection{Production from inflaton decay}

DM can also be produced during reheating by the direct decay of the inflaton. When the decay rate for {\em both} the dominant decay products of $\Phi$ and the DM particle is proportional to $m_{\Phi}$, the production rate in the Boltzmann equation (\ref{eq:DMBoltzmann}) takes the form
\begin{align}
R^d(T) \;&=\; \frac{y^2}{8\pi} B_{R} \rho_{\Phi}(T) \nonumber \\
&\simeq\; \frac{y^2}{8\pi}B_{R} \left(\frac{\pi^2 g_* T_{\rm reh}^4}{30}\right) \left( \frac{T}{T_{\rm reh}}\right)^{\frac{4k}{k-1}}\,, 
\end{align}
where $B_{R}$ denotes the branching ratio of the decay of the inflaton into DM, and includes the multiplicity of DM particles in the final state.
After a straightforward integration we obtain the following expression for the DM number density originating directly from inflaton decay\footnote{This result is modified if the decay rate of the inflaton to DM has a different dependence on $m_{\Phi}$. The corresponding DM number density for $\Gamma_{\Phi\rightarrow {\rm DM}}\propto m_{\Phi}^{\alpha+1}$ for generic $\alpha$ is 
left for future work.
},
\begin{align}
n^d (T_{\rm reh}) \;=\; &\frac{\sqrt{10g_*}}{480} (k+2) B_{R} y^2 M_P T_{\rm reh}^2  \,,
\label{ndec}
\end{align}
for low $m_{\rm DM}$ (the crossover mass is defined shortly).
For high $m_{\rm DM}$, 
we must cut off the integration at a temperature at which $m_\Phi = m_{\rm DM}$.
Recall that for $k > 2$, $m_{\Phi}$ evolves with $\Phi$. We can use Eq.~(\ref{Eq:mphi})
to determine, the temperature $T_L$, such that $m_\Phi(T_L) = m_{\rm DM}$ and find
\beq
\left( \frac{T_L}{T_{\rm reh}} \right)^\frac{4-2k}{k-1} = \frac{16\pi}{y^2} \sqrt{\frac{\pi^2 g_*}{90}} \left(\frac{7-k}{k+2} \right) \frac{T_{\rm reh}^2}{m_{\rm DM}{M_P} }\, .
\label{tl}
\eeq
In this case, 
\begin{align}
n^d (T_{\rm reh}) \;=\; &\frac{\sqrt{10g_*}}{480} (k+2) B_{R} y^2 M_P \left( \frac{T_{{\rm reh}}}{T_L} \right)^{\frac{4}{k-1}} T_{\rm reh}^2  \,.
\label{ndec2}
\end{align}
The crossover from Eq.~(\ref{ndec}) to Eq.~(\ref{ndec2}) occurs when $T_L = T_{\rm reh}$ and is easily obtained from Eq.~(\ref{tl}). We call the mass at the crossover $m_L$,
and 
\beq
m_L = \frac{16\pi}{y^2} \sqrt{\frac{\pi^2 g_*}{90}} \left(\frac{7-k}{k+2} \right) \frac{T_{\rm reh}^2}{{M_P} }\, .
\eeq

\subsection{The total dark matter relic abundance}

We next combine the dark matter densities produced by scattering and inflaton decay to obtain the following expression for the total present-day relic abundance,
\begin{figure*}[ht]
\centering
    \subfloat[]{\includegraphics[width=0.48\linewidth]{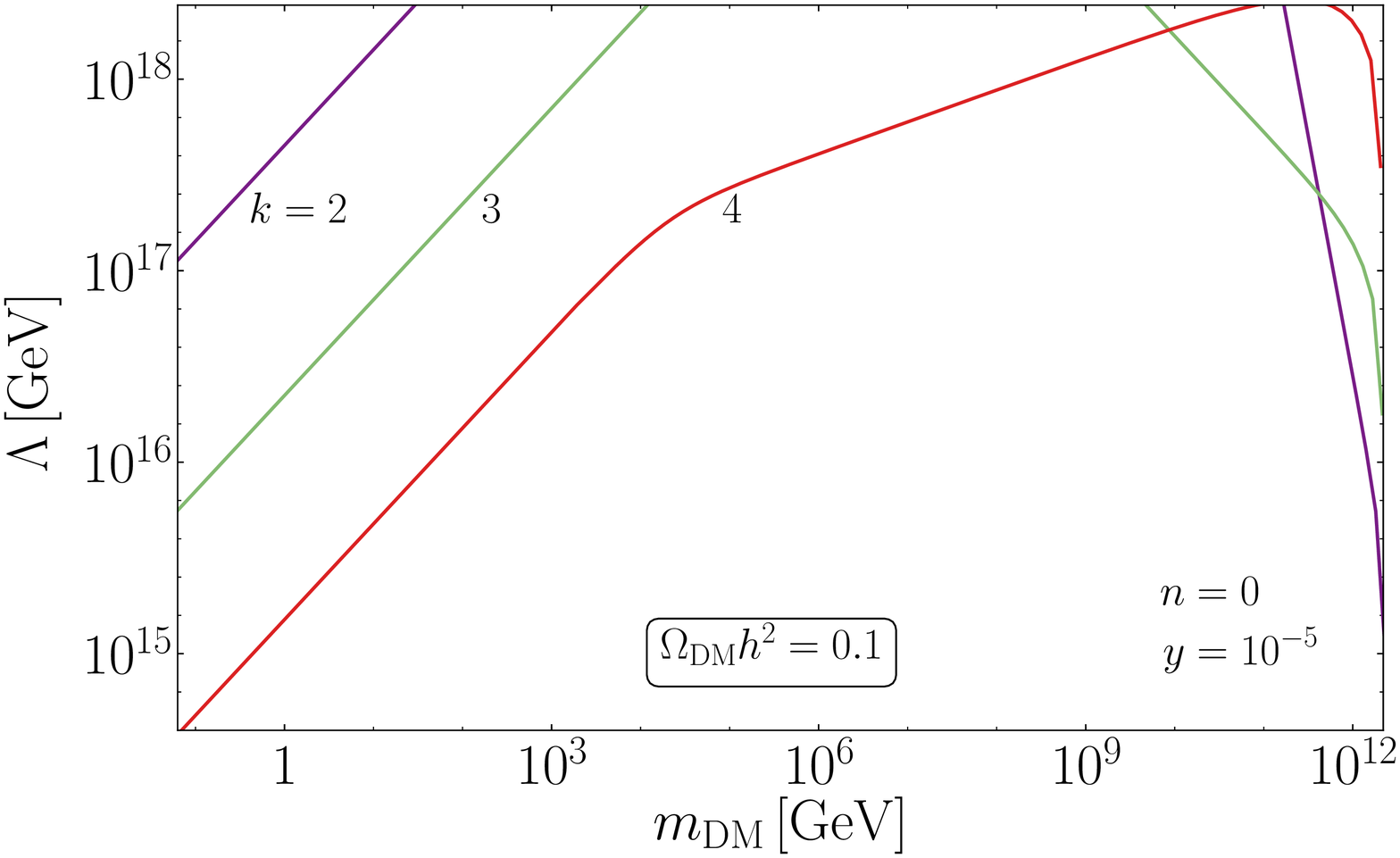}\label{sub:a}}
\hfill
    \subfloat[]{\includegraphics[width=0.48\linewidth]{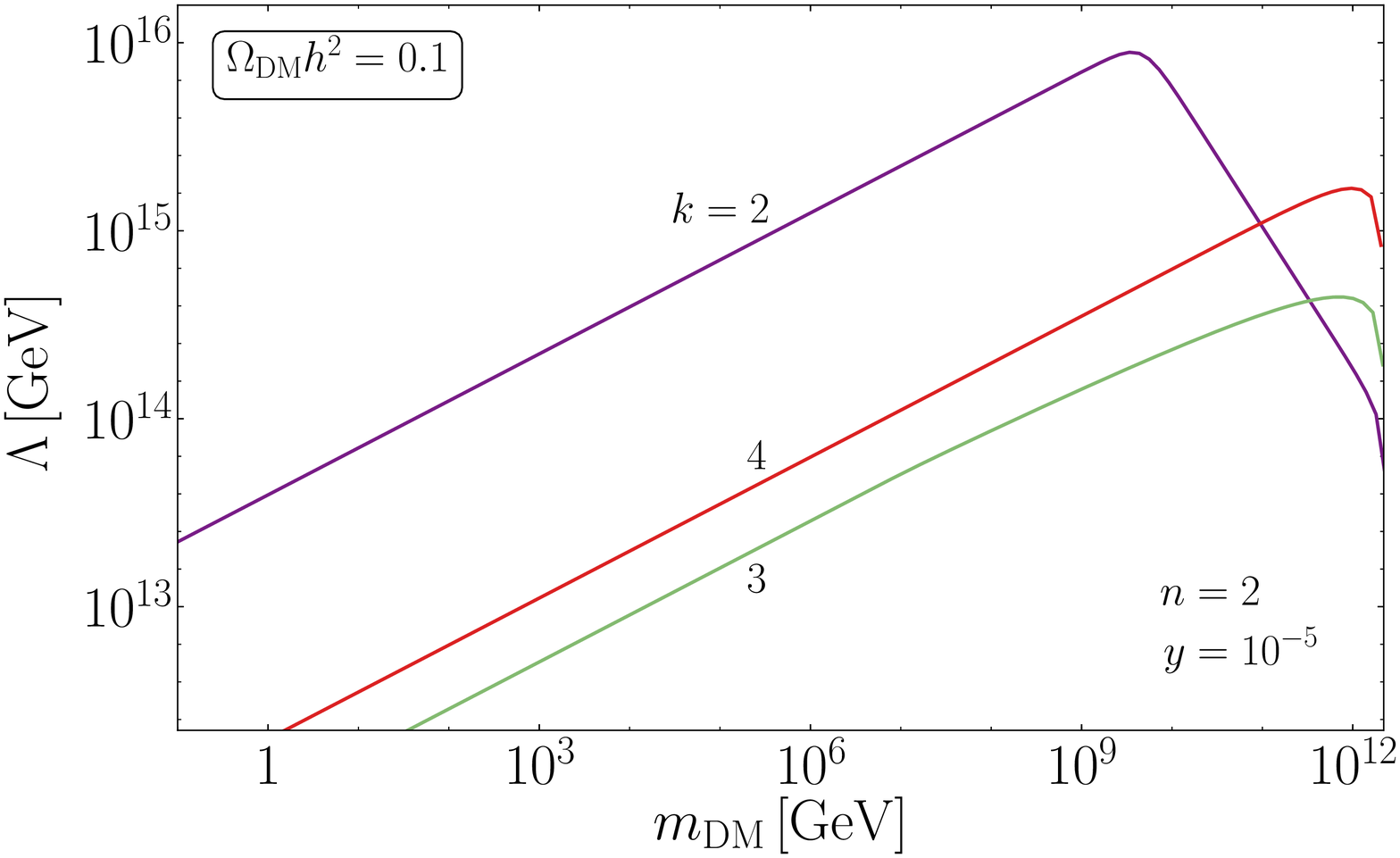}\label{sub:b}}

    \subfloat[]{\includegraphics[width=0.48\linewidth]{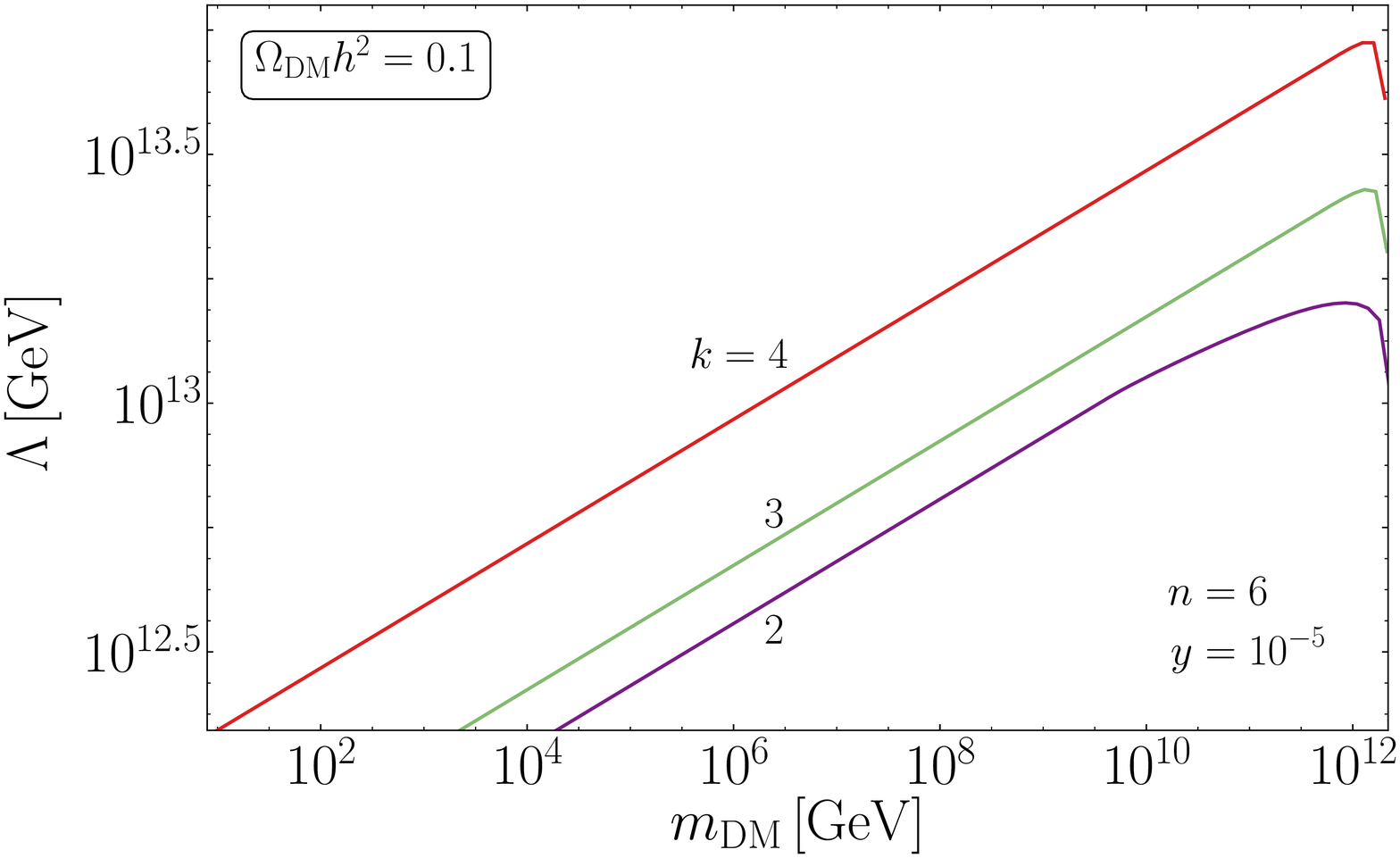}\label{sub:c}}
\hfill
    \subfloat[]{\includegraphics[width=0.48\linewidth]{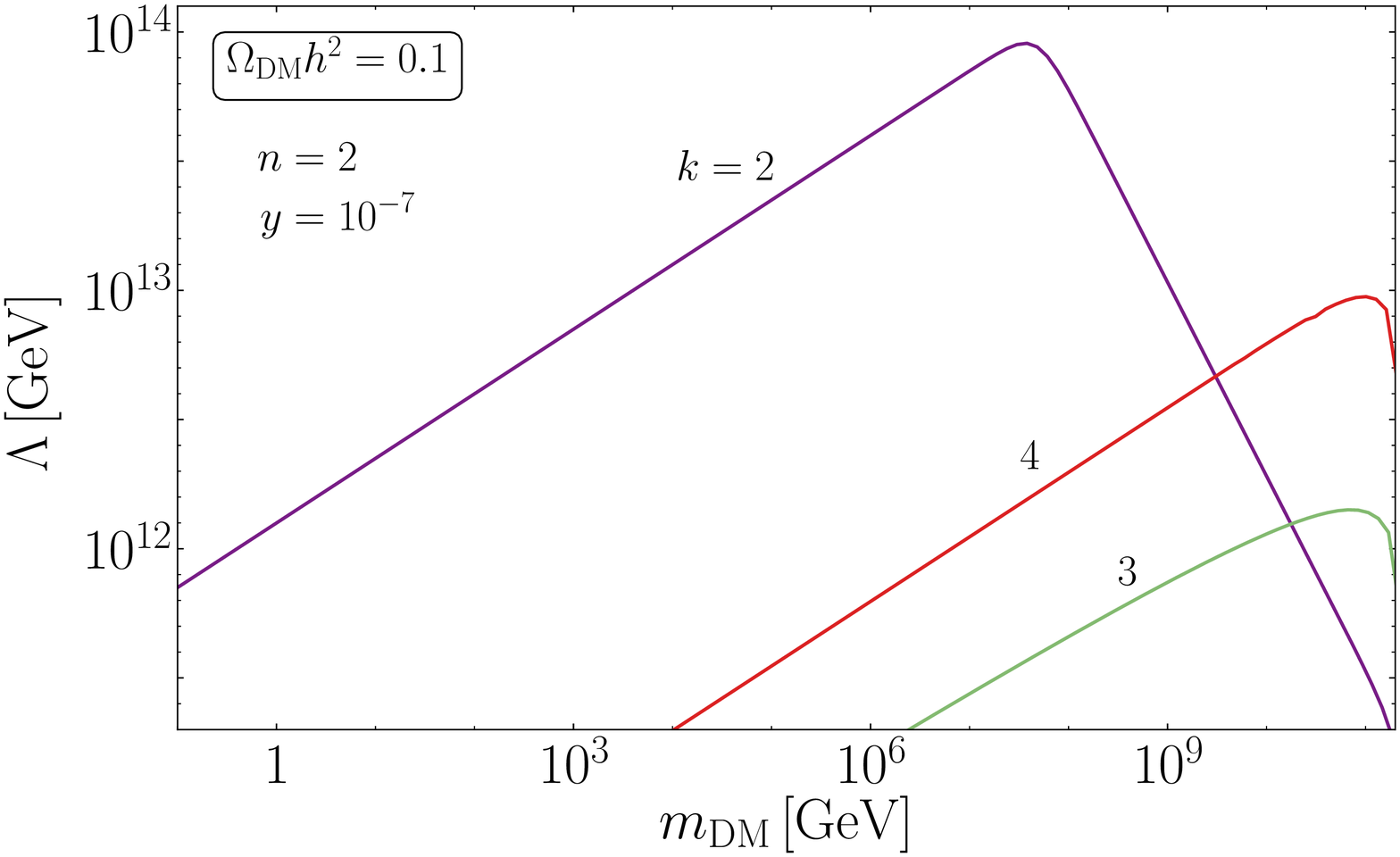}\label{sub:d}}
\caption{Contours $\Omega_{\rm DM}h^2=0.1$ showing the required value of $\Lambda$ as a function of the DM mass. We assume an inflaton decay coupling $y=10^{-5}$ and a production rate with $n=0$ (a), $n=2$ (b), and $n=6$ (c). For (d), we assume $y=10^{-7}$ and $n=2$. In all cases we have set $ B_R=0$.}
\label{fig:GenFern}
\end{figure*}
\begin{widetext}
\begin{align} \notag
\Omega_{\rm DM}h^2 \;\simeq\; &0.1  \times  \Bigg[ (k+2) \left(\frac{ B_{R}}{10^{-5}}\right) \left(\frac{427/4}{g_{\rm reh}}\right)^{1/2} \left(\frac{y}{10^{-5}}\right)^2 \left(\frac{10^{10}\,{\rm GeV}}{T_{\rm reh}}\right) \left(\frac{m_{\rm DM}}{1\,{\rm GeV}}\right) \nonumber \\
& \times \begin{cases}
1\,, \qquad &  m_{\rm DM} < m_{L}\, , \\
\left( \dfrac{T_{\rm reh}}{T_L} \right)^{\frac{4}{k-1}}\,, 
\qquad & m_{\rm DM} > m_{L}\, ,
\end{cases} \nonumber \\
& + 1.4 \times 10^{3-6n} \left(\dfrac{10^{16}\,{\rm GeV}}{\Lambda}\right)^{n+2}
\left(\dfrac{T_{\rm reh}}{10^{10}\,{\rm GeV}}\right)^{n+1} \left(\dfrac{427/4}{g_{\rm reh}}\right)^{3/2} \left(\dfrac{m_{\rm DM}}{1 \,{\rm GeV}}\right) \nonumber \\
& \times \begin{cases}
\dfrac{2k+4}{n-nk+10-2k}  \,,\quad & n<\dfrac{10-2k}{k-1} \qquad m_{\rm DM} < T_{\rm reh} \,,\\[10pt]
\dfrac{2k+4}{n-nk+10-2k} \left(\dfrac{T_{\rm reh}}{m_{\rm DM}} \right)^{\frac{2k+6}{k-1}-n-4}  \,,\quad & n<\dfrac{10-2k}{k-1} \qquad m_{\rm DM} > T_{\rm reh} \,,\\[10pt]
\dfrac{2k+4}{k-1}  \ln\left( \dfrac{T_{\rm max}}{T_{\rm reh}}\right) \,,\quad & n=\dfrac{10-2k}{k-1} \qquad  m_{\rm DM} < T_{\rm reh} \,,\\[10pt]
\dfrac{2k+4}{k-1}  \ln\left( \dfrac{T_{\rm max}}{m_{\rm DM}}\right) \,,\quad & n=\dfrac{10-2k}{k-1} \qquad  m_{\rm DM} > T_{\rm reh} \,,\\[10pt]
\dfrac{2k+4}{nk-n+2k-10} \left(\dfrac{T_{\rm max}}{T_{\rm reh}}\right)^{\frac{2k-10}{k-1}+n}  \,,\ & n>\dfrac{10-2k}{k-1}
\end{cases}
\Bigg]\,,
\label{Eq:omega}
\end{align}
\end{widetext}
where the first term corresponds to the production from decays, while the second, $\Lambda$-dependent term corresponds to freeze-in production through scattering. For the former term, it is worth noting that for $k=4$, Eq.~(\ref{eq:TrehAn}) implies that $T_{\rm reh}\propto y^2$, and therefore the decay contribution does not depend on the reheating temperature. It depends only on the square of the ratio of the inflaton-DM and inflaton-SM couplings, encoded in $B_{R}$, and the DM mass.
In the case of scatterings, we see clearly here the enhancement in ($T_{\rm max}/T_{\rm reh}$) for $n> (10-2k)/(k-1)$.

In Fig.~\ref{fig:GenFern},
we display the value of $\Lambda$ (in Eq.~(\ref{eq:RTs})) as a function of the DM mass, $m_{\rm DM}$,
needed to obtain $\Omega^s_{\rm DM} h^2 = 0.1$ in Eq.~(\ref{oh2s}) for $k=2, 3, 4$. In Fig.~\ref{fig:GenFern}\subref{sub:a}, we have chosen $n=0$ which is characteristic of a production rate for gravitinos in supersymmetric models when $\Lambda \sim M_P$. In this figure, we have chosen $y = 10^{-5}$. According to Fig.~\ref{fig:TgPlotA}, this corresponds to a value of $T_{\rm max} \sim 10^{12}$ GeV and $T_{\rm reh} \sim 10^{10}$ GeV. For $k = 2$, one gets the expected result that the density of gravitinos accounts for the DM
when $m_{3/2} \sim 100$ GeV, for $\Lambda \sim M_P$.

 As discussed in the earlier sections, fixing the inflaton decay coupling, $y$, fixes the maximum and final reheating temperature depending on the value of $k$. The relic density depends on $T_{\rm reh}$ through $n^s(T_{\rm reh})$ as given in Eqs.~(\ref{nlt})-(\ref{ngt}).
But $n^s$ also depends on $\Lambda^{-(n+2)}$. In Fig.~\ref{fig:GenFern}\subref{sub:a}, for $n=0$, the density is given by Eq.~(\ref{nlt}) and we see from Eq.~(\ref{Eq:omega})
that $\Omega^s_{\rm DM}$ scales as $m_{\rm DM}/\Lambda^2$
which accounts for the slope in the figure.
We also see that
the required value of $\Lambda$ decreases with increasing $k$ to compensate for the lower reheat temperature when $k>2$. Suitable DM
masses range from 0.1 to $T_{\rm max}$ for $\Lambda = 10^{14} {\rm GeV}$ to $M_P$.

\begin{figure*}[ht]
\centering
    \subfloat[]{\includegraphics[width=0.48\linewidth]{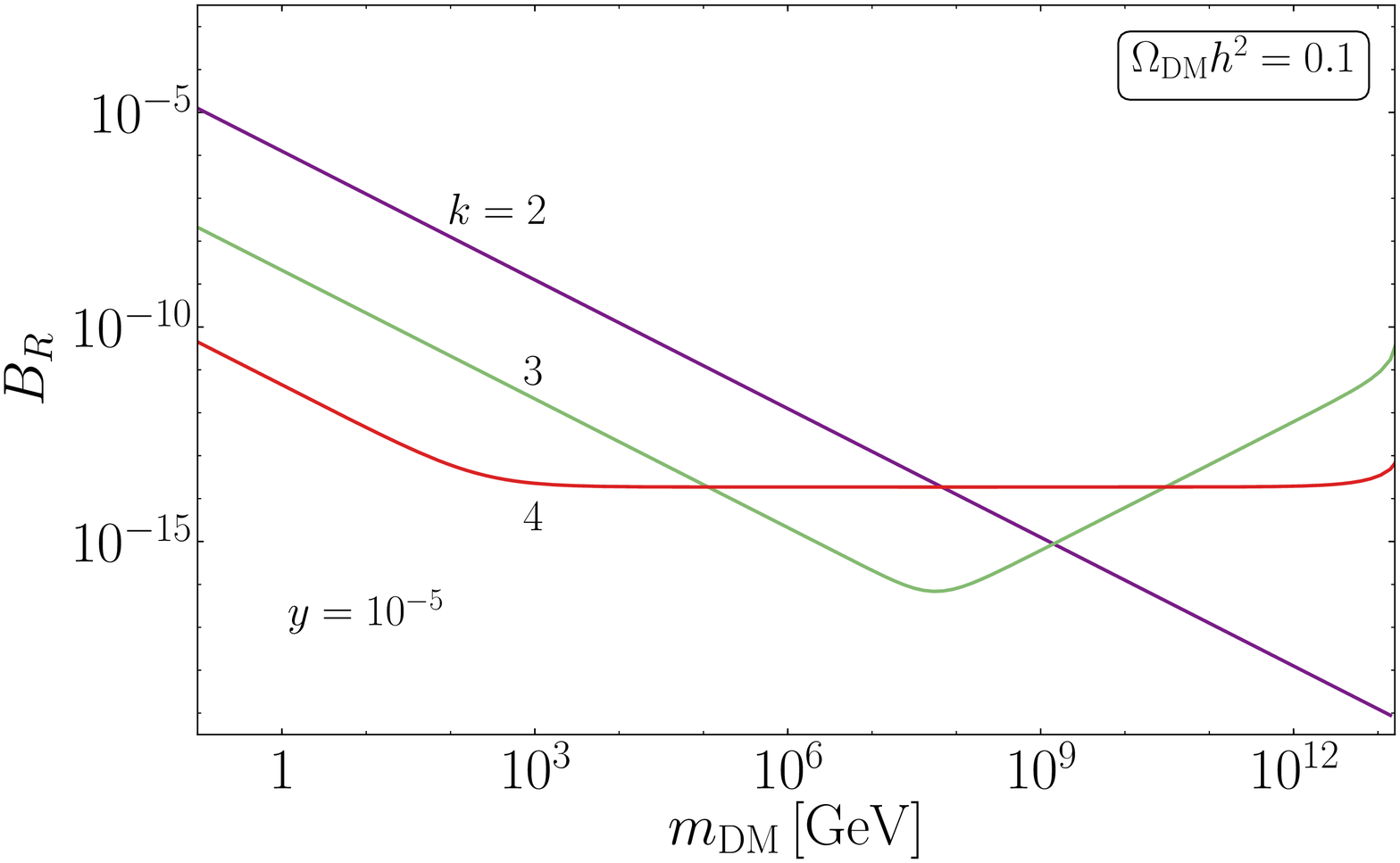}\label{sub:1}}
\hfill
    \subfloat[]{\includegraphics[width=0.48\linewidth]{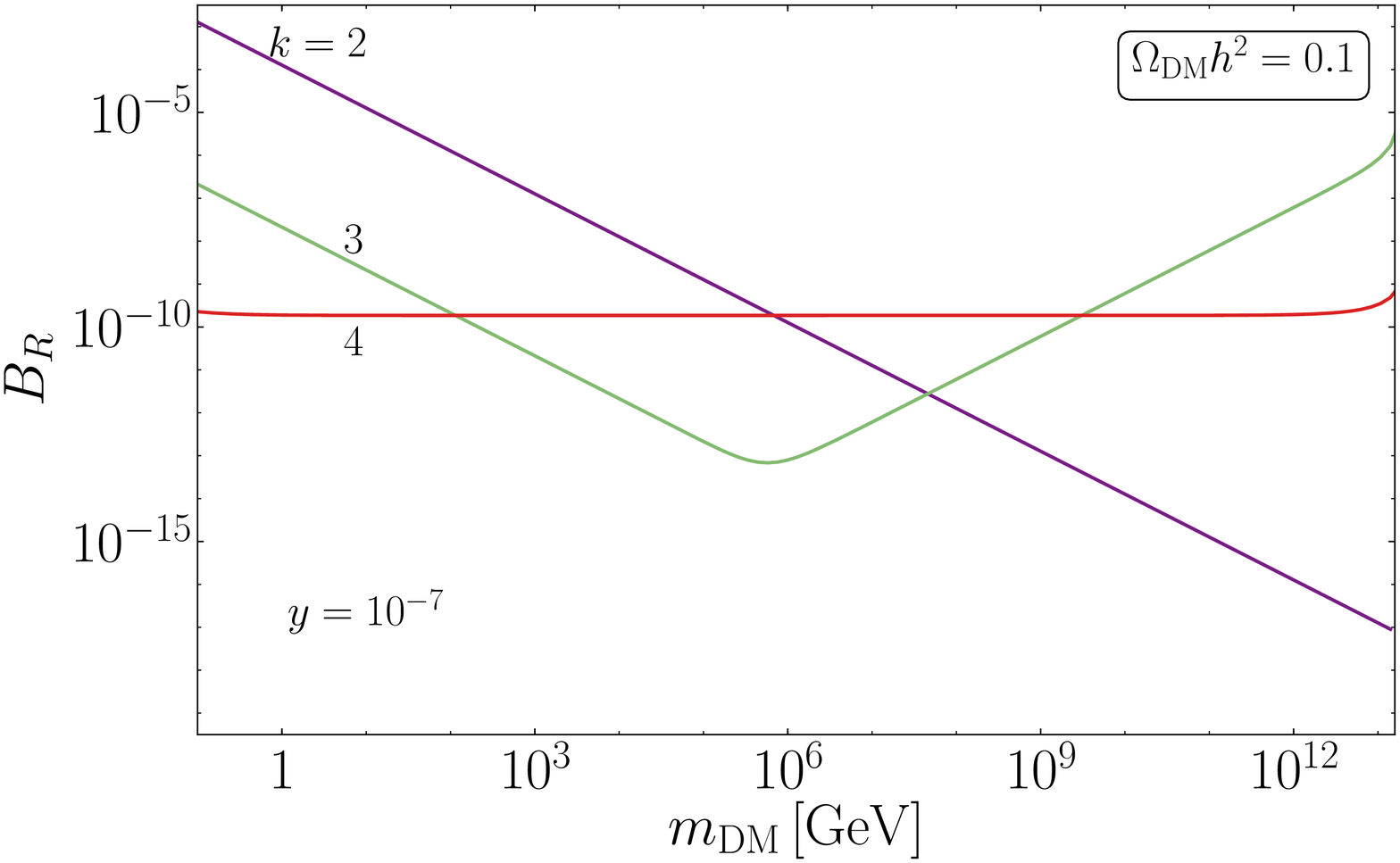}\label{sub:2}}
\caption{Contours $\Omega_{\rm DM}h^2=0.1$ showing the required branching ratio as a function of the DM mass, assuming an inflaton decay coupling $y=10^{-5}$ (a) and $y=10^{-7}$ (b). Here, we ignore production due to scattering.}
\label{fig:GenFern7}
\end{figure*}

In Fig.~\ref{fig:GenFern}\subref{sub:a} we also see changes in the slopes of the lines for all three values of $k$.
These occur when $m_{\rm DM} = T_{\rm reh}$ as discussed earlier. For $k = 2$ and 3, the change in slope occurs at
high $\Lambda (> M_P)$ and is off the scale of the plot.
The relative slope seen for $m_{\rm DM} > T_{\rm reh}$
can also be understood from Eq.~(\ref{Eq:omega}),
noting that $\Omega^s_{\rm DM}$ scales as $m_{\rm DM}^{\frac{9-k}{k-1}}/\Lambda^2$, so that for $k = 2$,
$\Lambda$ must decrease  steeply, whereas for $k=4$,
it remains an increasing function of $m_{\rm DM}$.  
All of the curves fall off when $m_{\rm DM}$ approaches $T_{\rm max}$.

In Fig.~\ref{fig:GenFern}\subref{sub:b}, we show the corresponding result when $n = 2$. 
Models with $n=2$ could correspond to interactions mediated by the exchange of a massive particle with mass $\Lambda > T_{\rm max}$. 
In this case, $k=3$ is the critical value, and the density becomes sensitive to $T_{\rm max}$
as in Eq.~(\ref{neq}). For larger $k$, the density is given
by Eq.~(\ref{ngt}) and exhibits a stronger dependence on $T_{\rm max}$.
For this reason, we see that a larger value of $\Lambda$ is required to obtain $\Omega^s_{\rm DM} h^2 = 0.1$ for $k =4$ than for $k=3$. 
Overall however, we see that lower values of $\Lambda$ are required for $n=2$ compared with $n=0$.

For $n=2$, we again see a change in slope for $k=2$ when 
$m_{\rm DM} = T_{\rm reh}$. In this case, $\Omega^s_{\rm DM}$ scales as $m_{\rm DM}^5/\Lambda^2$ at large DM masses.
For $k=3$ the change in slope is more subtle as
the dependence on $m_{\rm DM}$ goes from $m_{\rm DM}/\Lambda^2$ to $m_{\rm DM}\log(T_{\rm reh}/m_{\rm DM})/\Lambda^2$, and for $k=4$, there is no change as the density is primarily sensitive to $T_{\rm max}$.

In Fig.~\ref{fig:GenFern}\subref{sub:c}, we show the corresponding result when $n = 6$. Rates with $n=6$ could correspond to the production of gravitinos in high-scale supersymmetry \cite{Benakli:2017whb,grav2,grav3,grav4,Kaneta:2019zgw}
in which two gravitational vertices are required, or in models with spin-2 mediators \cite{Bernal:2018qlk}.
In the case of high-scale supersymmetry, we expect
that $\Lambda^2 \sim m_{3/2} M_P$. Thus, for $m_{3/2} \sim 1$ EeV, we should find $\Lambda \sim 10^{-5} M_P$ which is what is seen in Fig.~\ref{fig:GenFern}\subref{sub:c}. In this case, 
we are sensitive to $T_{\rm max}$ even for $k=2$ and the 
value of $\Lambda$ needed is now greater for $k=4$ than both $k=2$ and $k=3$. Note that there is again a subtle change in slope for $k=2$ when $m_{\rm DM}$ becomes larger than $T_{\rm reh}$ as was the case for $n=2$ and $k=3$. 

We can also use Eq.~(\ref{Eq:omega}) to compare predictions for the relic density for a given scattering rate (ignoring decays) for different values of $k$. Consider first the case with $n=2$. 
When comparing $k=2$ and $k=4$
for $n=2$, the large drop in $T_{\rm reh}^3$ when going from $k = 2$ to $k=4$ is not compensated by
the enhancement in $(T_{\rm max}/T_{\rm reh})^{4/3}$
and thus for a given value of $\Lambda$, we require
larger masses for $k=4$, as seen in Fig.~\ref{fig:GenFern}\subref{sub:b}.
In contrast when looking at the $n=6$ case,
the drop in $T_{\rm reh}^7$ is more than compensated for by the enhancement in $(T_{\rm max}/T_{\rm reh})^{16/3}$ and in this case for a given value of $\Lambda$, we require
smaller masses for $k=4$, as seen in Fig.~\ref{fig:GenFern}\subref{sub:c}.

To see the dependence of these results on $y$,
we show in  Fig.~\ref{fig:GenFern}\subref{sub:d} analogous results
with $y = 10^{-7}$ for $n = 2$. Since the reheating temperature is lower (for lower $y$), the necessary value of $\Lambda$ is also lower. Once again we see a change in slope for $k=2$
at $m_{\rm DM} = T_{\rm reh}$, the logarithmic change for $k=3$, and no change in slope for $k=4$, as in Fig.~\ref{fig:GenFern}\subref{sub:b}. 

In Fig.~\ref{fig:GenFern7}\subref{sub:1}, we show the necessary branching ratio to obtain $\Omega_{\rm DM} h^2 = 0.1$ from decay, assuming $y = 10^{-5}$, using Eqs.~(\ref{ndec}) and (\ref{ndec2})
and ignoring any possible contribution from scattering.
Since $\Omega_{\rm DM} h^2 \propto T_{\rm reh}^{-1}$ (see Eq.~(\ref{Eq:omega}) above),
and since $T_{\rm reh}$ drops significantly with increasing $k$, 
$B_{R}$ must be smaller for larger $k$,
as seen in the figure.
For a given value of $k$, $\Omega_{\rm DM} \propto B_{R} m_{\rm DM}$, accounting for the slope
in the lines shown in the figure. 

This behavior is strictly true in the whole kinematically allowed range only for $k=2$, or for a relatively small $m_{\rm DM}$ for $k>2$. As discussed earlier, this is because for $k>2$, when $m_{\rm DM} > m_L$, decays
to DM cannot continue all the way down to $T_{\rm reh}$. As a consequence the slope of the required branching ratio $B_R$ vs. $m_{\rm DM}$ changes.  From Eqs.~(\ref{tl}) and (\ref{ndec2}), we see that $T_L \propto m_{\rm DM}^{(1-k)/(4-2k)}$ and hence $\Omega^d_{\rm DM} h^2 \propto B_R n^d m_{\rm DM} \propto B_R T_L^{4/(1-k)} m_{\rm DM} \propto B_R m^{(8-2k)/(4-2k)}_{\rm DM}$. For $k=3$,  this gives
$\Omega^d_{\rm DM} h^2 \propto B_R/m_{\rm DM}$ and for $k=4$, $\Omega^d_{\rm DM} h^2 \propto B_R$, independent of $m_{\rm DM}$,
thus explaining the slopes seen in the figure.

 In Fig.~\ref{fig:GenFern7}\subref{sub:2}, we show the dependence of $B_{\rm DM}$ on $m_{\rm DM}$ for $y = 10^{-7}$.
 Note that there is some dependence on $y$ in $T_{\rm reh} \sim y^{k/2}$
 as seen in Eq.~(\ref{eq:TrehAn}).  Therefore, $\Omega_{\rm DM} h^2 \propto B_{\rm DM} y^{2-k/2}$. 
 For $k=2$, we see that the required branching ratio is larger compared with that in Fig.~\ref{fig:GenFern7}. However, for $k=4$, the dependence on $y$ drops out, and the required branching ratio is unchanged. In this case, the crossover mass, $m_L$ is lower, and for $k=4$, it is below 1 GeV.

\section{V. Conclusion and discussion}

While inflation was designed to resolve a host of cosmological problems, such as flatness and isotropy, it
also seeds the fluctuations necessary for structure in the 
Universe. Also needed to form structure is the existence
of dark matter. If dark matter interactions with Standard
Model particles are so weak that they never attain thermal equilibrium with the Standard Model bath, their existence
may also be a result of an early inflationary period. 
More precisely, the origin of dark matter may reside 
in process of reheating after inflation. 

A classic example of dark matter born out of reheating 
is the gravitino. With Planck-suppressed couplings, 
the abundance of gravitinos is proportional to the reheating temperature after inflation \cite{ehnos}, though there may also be a non-thermal component from decays of either the inflaton, or the next lightest supersymmetric particle. 
Dark matter production during reheating may be the dominant 
production mechanism for a class of candidates known as FIMPs \cite{fimp1}.

While certain quantitative aspects of dark matter production can be ascertained from the instantaneous reheating approximation, the dark matter abundance may be grossly
underestimated if the dark matter production rate has
a strong temperature dependence  
as in the case of the gravitino in high-scale supersymmetric models \cite{Benakli:2017whb} or dark matter interactions 
mediated by spin-2 particles \cite{Bernal:2018qlk}.
Instantaneous reheating refers to the approximation
that all inflatons decay in an instant (usually defined to be the time when the inflaton decay rate,
$\Gamma_{\Phi} \simeq H$). At the same moment, the Universe
becomes radiation dominated with a temperature determined by the energy density stored in the inflaton at the time of decay. 

However, inflaton decay is never instantaneous \cite{Giudice:2000ex,Chung:1998rq,Garcia:2017tuj,grav2}.
If the inflaton decay products rapidly thermalize, a radiation bath is formed even though $\rho_R \ll \rho_\Phi$.
Depending on the coupling of dark matter to the Standard Model, dark matter production may begin soon after the first
decays occur. Indeed, the Universe will first heat up to a 
temperature $T_{\rm max}\gg T_{\rm reh}$, and the maximum temperature may ultimately determine the dark matter abundance. 

Inflation occurs in the part of field space where the scalar potential is relatively flat. The exit from the inflationary phase occurs as the inflaton settles to its minimum and begins the reheating process. Often it is assumed that 
the potential during reheating can be approximated by
a quadratic potential. 
In this paper, we studied the reheating process in the case of a generic 
inflaton potential which can be expressed as $V(\Phi) = \lambda \frac{|\Phi|^k}{M_P^{k-4}}$ about its minimum.
For $k \ne 2$, the Universe does not expand as if it were dominated by matter, rather it is subject to an equation of state given by $w = (k-2)/(k+2)$ \cite{Lozanov:2016hid,Bernal:2019mhf}. 
Here, 
we  have shown that the presence of an effective mass $m_\Phi(t)$ affects strongly
the evolution of the temperature, especially near the end of the inflation
where the reheating temperature $T_{\rm reh}$ is highly dependent on $k$ and can be significantly smaller than in the vanilla quadratic case $k=2$. 

We have parametrized the dark matter production rate
as $R \propto T^{n+6}$.  In the case, of $k=2$,
for $n<6$, the dark matter abundance is primarily determined by the reheating temperature. For $n=6$, the abundance is enhanced by $\log (T_{\rm max}/T_{\rm reh})$ and for $n>6$,
it is primarily determined by $T_{\rm max}$ \cite{Garcia:2017tuj}. This picture changes, however, when $k > 2$. The critical value of $n$  decreases with increasing $k$, and for $k=4$, the dark matter abundance
is sensitive to $T_{\rm max}$ for $n\ge 1$. In these
cases, it is also possible to produce dark matter with masses in excess of the reheating temperature. For completeness, we have also considered the effect of 
the equation of state on the dark matter abundance 
originating directly from inflaton decays. 

In this paper, we have focused on the effects of the equation of state during reheating. We have limited our
discussion to inflaton decays to fermions, neglecting thermal effects, and assumed that 
the decay width of the inflaton is simply proportional to the inflaton mass.  Both assumptions affect our quantitative results, and these will treated more generally in future work. We have also neglected the delay of the onset of thermal equilibrium and the self-interaction of the inflaton, which we will also consider in future work.
The scenario outlined in this paper also does not exhaust all DM production channels. DM can also be produced by the decay of the heavy particles that may be produced at the early stages of reheating, or (in)directly by non-adiabatic particle production~\cite{Kofman:1997yn,Greene:1997fu,Felder:1998vq,Amin:2014eta,Lozanov:2017hjm}. We also leave the study of these scenarios in a generic reheating stage for future work.

\vskip.1in
{\bf Acknowledgments:}
\noindent 
The authors want to thank especially Christophe Kulikowski for very  insightful discussions. This work was supported by the France-US PICS MicroDark. The work of Marcos Garcia was supported by the Spanish Agencia
Estatal de Investigaci\'on through Grants No.~FPA2015-65929-P (MINECO/FEDER, UE), PGC2018095161-B-I00, IFT Centro de Excelencia Severo
Ochoa SEV-2016-0597, and Red Consolider MultiDark
FPA2017-90566-REDC. 
Marcos Garcia and Kunio Kaneta acknowledge support by Institut Pascal at Universit\'e
Paris-Saclay during the Paris-Saclay Particle Symposium 2019, with the support of the P2I and SPU research departments and the P2IO Laboratory
of Excellence (program ``Investissements d'avenir''
ANR-11-IDEX-0003-01 Paris-Saclay and ANR-10-LABX-0038), as well as the IPhT. This project has received funding/support from the European Unions Horizon 2020 research and innovation programme under the Marie Skodowska-Curie grant agreements Elusives ITN No. 674896
and InvisiblesPlus RISE No. 690575. The work of Kunio Kaneta and Keith A. Olive was supported in part by the DOE Grant No.~${\rm DE\text{-}SC0011842}$ at the University of Minnesota.

\section*{Appendix}

\section{Inflationary modeling}

\subsection{T-attractors and supergravity}

The most recent measurements of the tilt of the primordial scalar power spectrum $n_s$ and the null detection of primordial tensor modes by the Planck Collaboration~\cite{planck3,Ade:2018gkx} appear to favor plateau-like potentials, characterized by relatively low energy densities. In this light, a lot of interest has been focused on a  class of models, which include the Starobinsky model \cite{Staro} and converge in their predictions to the ``attractor point''~\cite{Bezrukov:2007ep,Ellis:2013nxa,Kallosh:2013hoa,Kallosh:2013daa,Kallosh:2013yoa,Ferrara:2013rsa,Kallosh:2013maa,Kallosh:2013xya}
\beq
n_s \;\simeq\; 1- \frac{2}{N_*}\,, \qquad r\;\simeq\; \frac{12}{N_*^2}\,.
\label{eq:nsr}
\eeq
Here $r$ denotes the tensor-to-scalar ratio, and $N_*$ is the number of $e$-folds between the horizon crossing of the pivot scale $k_*$ and the end of inflation. 

Many of these models can be constructed \cite{ENO6} from no-scale supergravity~\cite{no-scale1,no-scale2,LN} defined by a K\"ahler potential of the form
\begin{equation}
\label{kah3}
K \; = \; -3  \ln\left(T + \overline{T} - \frac{|\phi|^2}{3}\right) \, ,
\end{equation}
where $T$ is a volume modulus and $\phi$ is a matter like field. Depending on the form of the superpotential,
either $T$ or $\phi$ can play the role of the inflaton \cite{Ellis:2013nxa,Ellis:2018zya}.
For example, the Starobinsky model is derived
from a simple Wess-Zumino-like superpotential \cite{ENO6},
\beq
W = M \left(\frac{\phi^2}{2} - \frac{\phi^3}{3\sqrt{3}} \right) \, ,
\eeq
where $\Phi$ is related to the canonically normalized inflaton through
\begin{equation}
\phi \;=\; \sqrt{3} \tanh\left(\frac{\Phi}{\sqrt{6}}\right)\,,
\label{kinphi}
\end{equation}
yielding the scalar potential
\begin{equation}
V = \left(\sqrt{3} M \frac{\tanh(\Phi/\sqrt{6})}{1+\tanh{(\Phi/\sqrt{6})}} \right)^2  =  \frac{3}{4} M^2 \left(1 - e^{- \sqrt{\frac{2}{3}} \Phi} \right)^{2},
\label{astaro}
\end{equation}
when $\langle T \rangle  = 1/2$. Here the inflaton mass, $M$
is fixed in a similar manner as is $\lambda$ from the CMB normalization as discussed below. 

Alternatively, if $\langle \phi \rangle = 0$ is fixed,
the superpotential \cite{Cecotti}
\beq
W = \sqrt{3} M \phi \left(T- \frac12\right) \, ,
\eeq
yields the same Starobinsky potential (\ref{astaro})
when
\beq
T = \frac12 e^{\sqrt{\frac23}\Phi}\, .
\eeq

A similar class of models sharing the attractor points in (\ref{eq:nsr}) can be derived from a superpotential of the form
\beq
W = 2^{\frac{k}{4}+1}\sqrt{\lambda}  \left( \frac{\phi^{\frac{k}{2}+1}}{k+2} - \frac{\phi^{\frac{k}{2}+3}}{3(k+6)} \right)\, .
\eeq
The resulting scalar potential is then
\beq\label{eq:Vattractor}
V(\Phi) \;=\; \lambda \left[ \sqrt{6} \tanh\left(\frac{\Phi}{\sqrt{6} }\right)\right]^k\,.
\eeq

Alternatively, choosing
\beq
W = \sqrt{\lambda}~\phi~(2T) \left(\sqrt{6} ~ \frac{2T-1}{2T+1} \right)^{\frac{k}{2}}\, ,
\eeq
yields the same potential given in Eq.~(\ref{eq:Vattractor})
and both
provide Planck-compatible completions for our potential (\ref{eq:Vgen}) at large field values~\cite{Kallosh:2013hoa}. In all of the above expressions, we have worked in units of $M_P$.
In the remainder of this Appendix, we will restore powers of $M_P$. \footnote{We note that for $k=3$, this formulation does not lead to a stable minimum at $\Phi = 0$.}

\subsection{Normalization of the potential}

In order to determine $\rho_{\rm end}$, one must find the inflaton field value at the end of inflation, defined where $\ddot{a}=0$ or equivalently $\dot{\Phi}_{\rm end}^2 = V(\Phi)$~\cite{Ellis:2015pla}. An approximate solution for this condition yields using (\ref{eq:Vattractor})
\beq
\Phi_{\rm end} \;\simeq\; \sqrt{\frac{3}{8}} M_P \ln\left[ \frac{1}{2}+\frac{k}{3}\left( k+\sqrt{k^2 +3} \right) \right]\,.
\label{Eq:phiend}
\eeq
For $k= (2,3,4)$, this yields $\Phi_{\rm end} = (0.78, 1.19, 1.50) M_P$, respectively, which can be compared with the Starobinsky result, $\Phi_{\rm end} = 0.62 M_P$ \cite{Ellis:2015pla}. Recall in addition that $\rho_{\rm end} = \frac32 V(\Phi_{\rm end})$ so that $\rho_{\rm end}/\lambda M_P^4  = (0.86,2.0,4.8)$ for $k= (2,3,4)$.

On the inflationary plateau, a series expansion of the inflationary potential allows us to relate the number of $e$-folds with the potential and its derivative. Namely, with~\cite{Kallosh:2013hoa} 
\beq
\frac{V(\Phi)}{\lambda M_P^4 6^{k/2}} \;=\;   1- 2k e^{-\sqrt{\frac{2}{3}}\frac{\Phi}{M_P}} + \mathcal{O}\left( k^2 e^{-2\sqrt{\frac{2}{3}} \frac{\Phi}{M_P}}\right) \,,
\eeq
the number of $e$-folds in the slow-roll approximation can be computed as
\beq
N_* \;\simeq\; \frac{1}{M_P^2} \int_{\Phi_{\rm end}}^{\Phi_*} d\Phi\, \frac{V(\Phi)}{V'(\Phi)} \;\simeq\; \sqrt{\frac{3}{2}}\frac{V_*}{M_P V_*^{\prime}}\,.
\eeq
Substitution into the slow-roll expression for the amplitude of the curvature power spectrum
\beq
A_{S*} \;\simeq\; \frac{V_*^3}{12\pi^2 M_P^6 (V_*^{\prime})^2}\,,
\eeq
finally gives
\beq\label{eq:lambdaT}
\lambda \;\simeq\; \frac{18\pi^2A_{S*}}{6^{k/2}N_*^2}\,.
\eeq
At the Planck pivot scale, $k_*=0.05\,{\rm Mpc}^{-1}$, $\ln(10^{10} A_{S*})=3.044$~\cite{planck2,planck3}. The number of $e$-folds must in general be computed numerically, as it is determined by the duration of reheating, which in turn is determined by the energy density at the end of reheating, dependent on $N_*$.

\subsection{The number of $e$-folds}

Finally, we provide numerical values for the number of $e$-folds between the exit of the horizon of the Planck pivot scale $k_*$ and the end of inflation for T-attractor inflation. 

Assuming no entropy production between the end of reheating and the reentry to the horizon of the scale $k_*$ in the late-time radiation or matter-dominated Universe, the number of $e$-folds will depend on the energy scale of inflation and the duration of reheating, parametrized by the equation-of-state parameter. Concretely,~\cite{Martin:2010kz,Liddle:2003as}
\begin{align}\notag
N_* \;=\; &\ln \left[ \frac{1}{\sqrt{3}}\left(\frac{\pi^2}{30}\right)^{1/4}\left(\frac{43}{11}\right)^{1/3}\frac{T_0}{H_0} \right] - \ln\left( \frac{k_*}{a_0H_0}\right) \\ \notag
&+ \frac{1}{4} \ln\left( \frac{V_*^2}{M_P^4 \rho_{\rm end}}\right) + \frac{1-3w_{\rm int}}{12(1+w_{\rm int})}\ln \left(\frac{\rho_{\rm reh}}{\rho_{\rm end}}\right)\\ \label{eq:Nstar}
& - \frac{1}{12}\ln g_{\rm reh}\,,
\end{align}
where the present temperature and Hubble parameter, as determined from CMB observations, are given by $T_0=2.7255\,{\rm K}$ and $H_0=67.36\,{\rm km\,s}^{-1}{\rm Mpc}^{-1}$~\cite{planck2,Fixsen:2009ug}. The present scale factor is normalized to $a_0=1$. The $e$-fold average of the equation of state parameter during reheating is denoted by $w_{\rm int}$, and for our purposes we will approximate it by $w$ given by (\ref{eq:wk}). The energy density at the end of reheating is denoted by $\rho_{\rm reh}$. The value of the potential at horizon crossing is approximated as $V_*\simeq 6^{k/2}\lambda M_P^4$, with $\lambda$ given by (\ref{eq:lambdaT}).

\begin{figure}[t!]
{\includegraphics[width=1.0\columnwidth]{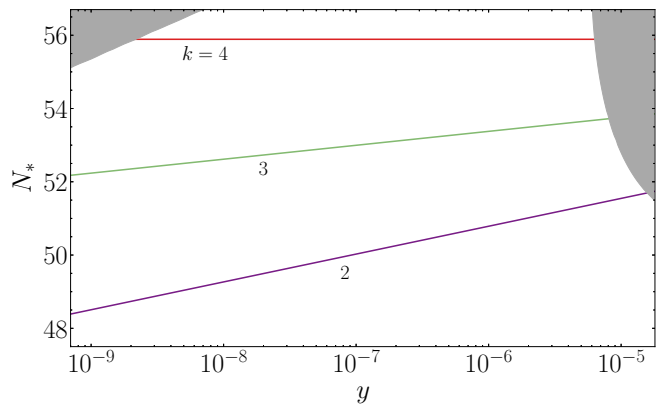}}
\caption{Number of $e$-folds from the exit of the Planck pivot scale $k_*=0.05\,{\rm Mpc}^{-1}$ to the end of inflation, as a function of $k$ and $y$. For definiteness we have used the Standard Model value $g_{\rm reh}=427/4$. In the shaded region our simple perturbative analysis breaks down (right), or $T_{\rm reh}<1\,{\rm MeV}$ (left).}
\label{fig:NsF}
\end{figure}

The numerical solution of Eq.~(\ref{eq:Nstar}) for the perturbative decay of the inflaton into fermions $f$ is shown in Fig.~\ref{fig:NsF} for $k=2,3,4$. The excluded region in gray corresponds either to the kinematic suppression regime (to the right), in which the effective masses satisfy the relation $m_f^2>m_{\Phi}^2$ at some point during reheating, and therefore where our analysis does not apply, or to reheating temperatures lower than $\sim 1\,{\rm MeV}$, incompatible with Big Bang Nucleosynthesis (to the left)~\cite{Hasegawa:2019jsa} (see also \cite{Fields:2019pfx}). Note that for all the values of $y$ shown, $N_* > 46$. Therefore, $n_s$ and $r$ here lie within the 95\% CL region of the Planck+BICEP2/Keck (PBK) constraints~\cite{Ade:2018gkx}. Moreover, for $N_* \gtrsim 50$, the model is compatible with PBK at the 68\% CL. Finally, we note that the curves become less steep for larger $k$. At $k=4$, $N_*\simeq 55.9$ independently of the decay rate, consistent with the fact that radiation domination (i.e.~$w=1/3$) is reached immediately after the end of inflation.


\bibliographystyle{apsrev4-1}

\bibliography{biblio} 

\begin{thebibliography}{96}%
\makeatletter
\providecommand \@ifxundefined [1]{%
 \@ifx{#1\undefined}
}%
\providecommand \@ifnum [1]{%
 \ifnum #1\expandafter \@firstoftwo
 \else \expandafter \@secondoftwo
 \fi
}%
\providecommand \@ifx [1]{%
 \ifx #1\expandafter \@firstoftwo
 \else \expandafter \@secondoftwo
 \fi
}%
\providecommand \natexlab [1]{#1}%
\providecommand \enquote  [1]{``#1''}%
\providecommand \bibnamefont  [1]{#1}%
\providecommand \bibfnamefont [1]{#1}%
\providecommand \citenamefont [1]{#1}%
\providecommand \href@noop [0]{\@secondoftwo}%
\providecommand \href [0]{\begingroup \@sanitize@url \@href}%
\providecommand \@href[1]{\@@startlink{#1}\@@href}%
\providecommand \@@href[1]{\endgroup#1\@@endlink}%
\providecommand \@sanitize@url [0]{\catcode `\\12\catcode `\$12\catcode
  `\&12\catcode `\#12\catcode `\^12\catcode `\_12\catcode `\%12\relax}%
\providecommand \@@startlink[1]{}%
\providecommand \@@endlink[0]{}%
\providecommand \url  [0]{\begingroup\@sanitize@url \@url }%
\providecommand \@url [1]{\endgroup\@href {#1}{\urlprefix }}%
\providecommand \urlprefix  [0]{URL }%
\providecommand \Eprint [0]{\href }%
\providecommand \doibase [0]{http://dx.doi.org/}%
\providecommand \selectlanguage [0]{\@gobble}%
\providecommand \bibinfo  [0]{\@secondoftwo}%
\providecommand \bibfield  [0]{\@secondoftwo}%
\providecommand \translation [1]{[#1]}%
\providecommand \BibitemOpen [0]{}%
\providecommand \bibitemStop [0]{}%
\providecommand \bibitemNoStop [0]{.\EOS\space}%
\providecommand \EOS [0]{\spacefactor3000\relax}%
\providecommand \BibitemShut  [1]{\csname bibitem#1\endcsname}%
\let\auto@bib@innerbib\@empty
\bibitem [{\citenamefont {Poincar\'e}(1906)}]{poincare}%
  \BibitemOpen
  \bibfield  {author} {\bibinfo {author} {\bibfnamefont {R.}~\bibnamefont
  {Poincar\'e}},\ }\href@noop {} {\bibfield  {journal} {\bibinfo  {journal}
  {L'Astronomie: revue mensuelle d'astronomie, de m\'et\'eorologie et de
  physique du globe et bulletin de la Soci\'et\'e astronomique de France}\ ,\
  \bibinfo {pages} {153}} (\bibinfo {year} {1906})}\BibitemShut {NoStop}%
\bibitem [{\citenamefont {Zwicky}(1933)}]{Zwicky:1933gu}%
  \BibitemOpen
  \bibfield  {author} {\bibinfo {author} {\bibfnamefont {F.}~\bibnamefont
  {Zwicky}},\ }\href {\doibase 10.1007/s10714-008-0707-4} {\bibfield  {journal}
  {\bibinfo  {journal} {Helv. Phys. Acta}\ }\textbf {\bibinfo {volume} {6}},\
  \bibinfo {pages} {110} (\bibinfo {year} {1933})},\ \bibinfo {note} {[Gen.
  Rel. Grav.41,207(2009)]}\BibitemShut {NoStop}%
\bibitem [{\citenamefont {Babcock}(1939)}]{babcock}%
  \BibitemOpen
  \bibfield  {author} {\bibinfo {author} {\bibfnamefont {H.~W.}\ \bibnamefont
  {Babcock}},\ }\href@noop {} {\bibfield  {journal} {\bibinfo  {journal} {Lick
  Observatory bulletin}\ }\textbf {\bibinfo {volume} {498}},\ \bibinfo {pages}
  {41} (\bibinfo {year} {1939})}\BibitemShut {NoStop}%
\bibitem [{\citenamefont {Gunn}\ \emph {et~al.}(1978)\citenamefont {Gunn},
  \citenamefont {Lee}, \citenamefont {Lerche}, \citenamefont {Schramm},\ and\
  \citenamefont {Steigman}}]{Gunn:1978gr}%
  \BibitemOpen
  \bibfield  {author} {\bibinfo {author} {\bibfnamefont {J.}~\bibnamefont
  {Gunn}}, \bibinfo {author} {\bibfnamefont {B.}~\bibnamefont {Lee}}, \bibinfo
  {author} {\bibfnamefont {I.}~\bibnamefont {Lerche}}, \bibinfo {author}
  {\bibfnamefont {D.}~\bibnamefont {Schramm}}, \ and\ \bibinfo {author}
  {\bibfnamefont {G.}~\bibnamefont {Steigman}},\ }\href {\doibase
  10.1086/156335} {\bibfield  {journal} {\bibinfo  {journal} {Astrophys.\ J.}\
  }\textbf {\bibinfo {volume} {223}},\ \bibinfo {pages} {1015} (\bibinfo {year}
  {1978})}\BibitemShut {NoStop}%
\bibitem [{\citenamefont {Aprile}\ \emph {et~al.}(2018)\citenamefont {Aprile}
  \emph {et~al.}}]{XENON}%
  \BibitemOpen
  \bibfield  {author} {\bibinfo {author} {\bibfnamefont {E.}~\bibnamefont
  {Aprile}} \emph {et~al.} (\bibinfo {collaboration} {XENON}),\ }\href
  {\doibase 10.1103/PhysRevLett.121.111302} {\bibfield  {journal} {\bibinfo
  {journal} {Phys. Rev. Lett.}\ }\textbf {\bibinfo {volume} {121}},\ \bibinfo
  {pages} {111302} (\bibinfo {year} {2018})},\ \Eprint
  {http://arxiv.org/abs/1805.12562} {arXiv:1805.12562 [astro-ph.CO]}
  \BibitemShut {NoStop}%
\bibitem [{\citenamefont {Akerib}\ \emph {et~al.}(2017)\citenamefont {Akerib}
  \emph {et~al.}}]{LUX}%
  \BibitemOpen
  \bibfield  {author} {\bibinfo {author} {\bibfnamefont {D.~S.}\ \bibnamefont
  {Akerib}} \emph {et~al.} (\bibinfo {collaboration} {LUX}),\ }\href {\doibase
  10.1103/PhysRevLett.118.021303} {\bibfield  {journal} {\bibinfo  {journal}
  {Phys. Rev. Lett.}\ }\textbf {\bibinfo {volume} {118}},\ \bibinfo {pages}
  {021303} (\bibinfo {year} {2017})},\ \Eprint
  {http://arxiv.org/abs/1608.07648} {arXiv:1608.07648 [astro-ph.CO]}
  \BibitemShut {NoStop}%
\bibitem [{\citenamefont {Cui}\ \emph {et~al.}(2017)\citenamefont {Cui} \emph
  {et~al.}}]{PANDAX}%
  \BibitemOpen
  \bibfield  {author} {\bibinfo {author} {\bibfnamefont {X.}~\bibnamefont
  {Cui}} \emph {et~al.} (\bibinfo {collaboration} {PandaX-II}),\ }\href
  {\doibase 10.1103/PhysRevLett.119.181302} {\bibfield  {journal} {\bibinfo
  {journal} {Phys. Rev. Lett.}\ }\textbf {\bibinfo {volume} {119}},\ \bibinfo
  {pages} {181302} (\bibinfo {year} {2017})},\ \Eprint
  {http://arxiv.org/abs/1708.06917} {arXiv:1708.06917 [astro-ph.CO]}
  \BibitemShut {NoStop}%
\bibitem [{\citenamefont {Arcadi}\ \emph {et~al.}(2018)\citenamefont {Arcadi},
  \citenamefont {Dutra}, \citenamefont {Ghosh}, \citenamefont {Lindner},
  \citenamefont {Mambrini}, \citenamefont {Pierre}, \citenamefont {Profumo},\
  and\ \citenamefont {Queiroz}}]{Arcadi:2017kky}%
  \BibitemOpen
  \bibfield  {author} {\bibinfo {author} {\bibfnamefont {G.}~\bibnamefont
  {Arcadi}}, \bibinfo {author} {\bibfnamefont {M.}~\bibnamefont {Dutra}},
  \bibinfo {author} {\bibfnamefont {P.}~\bibnamefont {Ghosh}}, \bibinfo
  {author} {\bibfnamefont {M.}~\bibnamefont {Lindner}}, \bibinfo {author}
  {\bibfnamefont {Y.}~\bibnamefont {Mambrini}}, \bibinfo {author}
  {\bibfnamefont {M.}~\bibnamefont {Pierre}}, \bibinfo {author} {\bibfnamefont
  {S.}~\bibnamefont {Profumo}}, \ and\ \bibinfo {author} {\bibfnamefont
  {F.~S.}\ \bibnamefont {Queiroz}},\ }\href {\doibase
  10.1140/epjc/s10052-018-5662-y} {\bibfield  {journal} {\bibinfo  {journal}
  {Eur. Phys. J.}\ }\textbf {\bibinfo {volume} {C78}},\ \bibinfo {pages} {203}
  (\bibinfo {year} {2018})},\ \Eprint {http://arxiv.org/abs/1703.07364}
  {arXiv:1703.07364 [hep-ph]} \BibitemShut {NoStop}%
\bibitem [{\citenamefont {Silveira}\ and\ \citenamefont {Zee}(1985)}]{hp1}%
  \BibitemOpen
  \bibfield  {author} {\bibinfo {author} {\bibfnamefont {V.}~\bibnamefont
  {Silveira}}\ and\ \bibinfo {author} {\bibfnamefont {A.}~\bibnamefont {Zee}},\
  }\href {\doibase 10.1016/0370-2693(85)90624-0} {\bibfield  {journal}
  {\bibinfo  {journal} {Phys. Lett.}\ }\textbf {\bibinfo {volume} {161B}},\
  \bibinfo {pages} {136} (\bibinfo {year} {1985})}\BibitemShut {NoStop}%
\bibitem [{\citenamefont {McDonald}(1994)}]{hp2}%
  \BibitemOpen
  \bibfield  {author} {\bibinfo {author} {\bibfnamefont {J.}~\bibnamefont
  {McDonald}},\ }\href {\doibase 10.1103/PhysRevD.50.3637} {\bibfield
  {journal} {\bibinfo  {journal} {Phys. Rev.}\ }\textbf {\bibinfo {volume}
  {D50}},\ \bibinfo {pages} {3637} (\bibinfo {year} {1994})},\ \Eprint
  {http://arxiv.org/abs/hep-ph/0702143} {arXiv:hep-ph/0702143 [HEP-PH]}
  \BibitemShut {NoStop}%
\bibitem [{\citenamefont {Burgess}\ \emph {et~al.}(2001)\citenamefont
  {Burgess}, \citenamefont {Pospelov},\ and\ \citenamefont {ter
  Veldhuis}}]{hp3}%
  \BibitemOpen
  \bibfield  {author} {\bibinfo {author} {\bibfnamefont {C.~P.}\ \bibnamefont
  {Burgess}}, \bibinfo {author} {\bibfnamefont {M.}~\bibnamefont {Pospelov}}, \
  and\ \bibinfo {author} {\bibfnamefont {T.}~\bibnamefont {ter Veldhuis}},\
  }\href {\doibase 10.1016/S0550-3213(01)00513-2} {\bibfield  {journal}
  {\bibinfo  {journal} {Nucl. Phys.}\ }\textbf {\bibinfo {volume} {B619}},\
  \bibinfo {pages} {709} (\bibinfo {year} {2001})},\ \Eprint
  {http://arxiv.org/abs/hep-ph/0011335} {arXiv:hep-ph/0011335 [hep-ph]}
  \BibitemShut {NoStop}%
\bibitem [{\citenamefont {Djouadi}\ \emph {et~al.}(2012)\citenamefont
  {Djouadi}, \citenamefont {Lebedev}, \citenamefont {Mambrini},\ and\
  \citenamefont {Quevillon}}]{Higgsportal2}%
  \BibitemOpen
  \bibfield  {author} {\bibinfo {author} {\bibfnamefont {A.}~\bibnamefont
  {Djouadi}}, \bibinfo {author} {\bibfnamefont {O.}~\bibnamefont {Lebedev}},
  \bibinfo {author} {\bibfnamefont {Y.}~\bibnamefont {Mambrini}}, \ and\
  \bibinfo {author} {\bibfnamefont {J.}~\bibnamefont {Quevillon}},\ }\href
  {\doibase 10.1016/j.physletb.2012.01.062} {\bibfield  {journal} {\bibinfo
  {journal} {Phys. Lett.}\ }\textbf {\bibinfo {volume} {B709}},\ \bibinfo
  {pages} {65} (\bibinfo {year} {2012})},\ \Eprint
  {http://arxiv.org/abs/1112.3299} {arXiv:1112.3299 [hep-ph]} \BibitemShut
  {NoStop}%
\bibitem [{\citenamefont {Djouadi}\ \emph {et~al.}(2013)\citenamefont
  {Djouadi}, \citenamefont {Falkowski}, \citenamefont {Mambrini},\ and\
  \citenamefont {Quevillon}}]{Higgsportal3}%
  \BibitemOpen
  \bibfield  {author} {\bibinfo {author} {\bibfnamefont {A.}~\bibnamefont
  {Djouadi}}, \bibinfo {author} {\bibfnamefont {A.}~\bibnamefont {Falkowski}},
  \bibinfo {author} {\bibfnamefont {Y.}~\bibnamefont {Mambrini}}, \ and\
  \bibinfo {author} {\bibfnamefont {J.}~\bibnamefont {Quevillon}},\ }\href
  {\doibase 10.1140/epjc/s10052-013-2455-1} {\bibfield  {journal} {\bibinfo
  {journal} {Eur. Phys. J.}\ }\textbf {\bibinfo {volume} {C73}},\ \bibinfo
  {pages} {2455} (\bibinfo {year} {2013})},\ \Eprint
  {http://arxiv.org/abs/1205.3169} {arXiv:1205.3169 [hep-ph]} \BibitemShut
  {NoStop}%
\bibitem [{\citenamefont {Mambrini}(2011)}]{Higgsportal5}%
  \BibitemOpen
  \bibfield  {author} {\bibinfo {author} {\bibfnamefont {Y.}~\bibnamefont
  {Mambrini}},\ }\href {\doibase 10.1103/PhysRevD.84.115017} {\bibfield
  {journal} {\bibinfo  {journal} {Phys. Rev.}\ }\textbf {\bibinfo {volume}
  {D84}},\ \bibinfo {pages} {115017} (\bibinfo {year} {2011})},\ \Eprint
  {http://arxiv.org/abs/1108.0671} {arXiv:1108.0671 [hep-ph]} \BibitemShut
  {NoStop}%
\bibitem [{\citenamefont {Alves}\ \emph {et~al.}(2014)\citenamefont {Alves},
  \citenamefont {Profumo},\ and\ \citenamefont {Queiroz}}]{Zpportal1}%
  \BibitemOpen
  \bibfield  {author} {\bibinfo {author} {\bibfnamefont {A.}~\bibnamefont
  {Alves}}, \bibinfo {author} {\bibfnamefont {S.}~\bibnamefont {Profumo}}, \
  and\ \bibinfo {author} {\bibfnamefont {F.~S.}\ \bibnamefont {Queiroz}},\
  }\href {\doibase 10.1007/JHEP04(2014)063} {\bibfield  {journal} {\bibinfo
  {journal} {JHEP}\ }\textbf {\bibinfo {volume} {04}},\ \bibinfo {pages} {063}
  (\bibinfo {year} {2014})},\ \Eprint {http://arxiv.org/abs/1312.5281}
  {arXiv:1312.5281 [hep-ph]} \BibitemShut {NoStop}%
\bibitem [{\citenamefont {Lebedev}\ and\ \citenamefont
  {Mambrini}(2014)}]{Zpportal2}%
  \BibitemOpen
  \bibfield  {author} {\bibinfo {author} {\bibfnamefont {O.}~\bibnamefont
  {Lebedev}}\ and\ \bibinfo {author} {\bibfnamefont {Y.}~\bibnamefont
  {Mambrini}},\ }\href {\doibase 10.1016/j.physletb.2014.05.025} {\bibfield
  {journal} {\bibinfo  {journal} {Phys. Lett.}\ }\textbf {\bibinfo {volume}
  {B734}},\ \bibinfo {pages} {350} (\bibinfo {year} {2014})},\ \Eprint
  {http://arxiv.org/abs/1403.4837} {arXiv:1403.4837 [hep-ph]} \BibitemShut
  {NoStop}%
\bibitem [{\citenamefont {Arcadi}\ \emph {et~al.}(2014)\citenamefont {Arcadi},
  \citenamefont {Mambrini}, \citenamefont {Tytgat},\ and\ \citenamefont
  {Zaldivar}}]{Zpportal3}%
  \BibitemOpen
  \bibfield  {author} {\bibinfo {author} {\bibfnamefont {G.}~\bibnamefont
  {Arcadi}}, \bibinfo {author} {\bibfnamefont {Y.}~\bibnamefont {Mambrini}},
  \bibinfo {author} {\bibfnamefont {M.~H.~G.}\ \bibnamefont {Tytgat}}, \ and\
  \bibinfo {author} {\bibfnamefont {B.}~\bibnamefont {Zaldivar}},\ }\href
  {\doibase 10.1007/JHEP03(2014)134} {\bibfield  {journal} {\bibinfo  {journal}
  {JHEP}\ }\textbf {\bibinfo {volume} {03}},\ \bibinfo {pages} {134} (\bibinfo
  {year} {2014})},\ \Eprint {http://arxiv.org/abs/1401.0221} {arXiv:1401.0221
  [hep-ph]} \BibitemShut {NoStop}%
\bibitem [{\citenamefont {Goldberg}(1983)}]{Go1983}%
  \BibitemOpen
  \bibfield  {author} {\bibinfo {author} {\bibfnamefont {H.}~\bibnamefont
  {Goldberg}},\ }\href {\doibase 10.1103/PhysRevLett.103.099905,
  10.1103/PhysRevLett.50.1419} {\bibfield  {journal} {\bibinfo  {journal}
  {Phys. Rev. Lett.}\ }\textbf {\bibinfo {volume} {50}},\ \bibinfo {pages}
  {1419} (\bibinfo {year} {1983})},\ \bibinfo {note} {[Erratum: Phys. Rev.
  Lett.103,099905(2009); ,219(1983)]}\BibitemShut {NoStop}%
\bibitem [{\citenamefont {Ellis}\ \emph
  {et~al.}(1984{\natexlab{a}})\citenamefont {Ellis}, \citenamefont {Hagelin},
  \citenamefont {Nanopoulos}, \citenamefont {Olive},\ and\ \citenamefont
  {Srednicki}}]{ehnos}%
  \BibitemOpen
  \bibfield  {author} {\bibinfo {author} {\bibfnamefont {J.~R.}\ \bibnamefont
  {Ellis}}, \bibinfo {author} {\bibfnamefont {J.~S.}\ \bibnamefont {Hagelin}},
  \bibinfo {author} {\bibfnamefont {D.~V.}\ \bibnamefont {Nanopoulos}},
  \bibinfo {author} {\bibfnamefont {K.~A.}\ \bibnamefont {Olive}}, \ and\
  \bibinfo {author} {\bibfnamefont {M.}~\bibnamefont {Srednicki}},\ }\href
  {\doibase 10.1016/0550-3213(84)90461-9} {\bibfield  {journal} {\bibinfo
  {journal} {Nucl. Phys.}\ }\textbf {\bibinfo {volume} {B238}},\ \bibinfo
  {pages} {453} (\bibinfo {year} {1984}{\natexlab{a}})},\ \bibinfo {note}
  {[,223(1983)]}\BibitemShut {NoStop}%
\bibitem [{\citenamefont {Bagnaschi}\ \emph {et~al.}(2019)\citenamefont
  {Bagnaschi} \emph {et~al.}}]{cmssm3}%
  \BibitemOpen
  \bibfield  {author} {\bibinfo {author} {\bibfnamefont {E.}~\bibnamefont
  {Bagnaschi}} \emph {et~al.},\ }\href {\doibase
  10.1140/epjc/s10052-019-6658-y} {\bibfield  {journal} {\bibinfo  {journal}
  {Eur. Phys. J.}\ }\textbf {\bibinfo {volume} {C79}},\ \bibinfo {pages} {149}
  (\bibinfo {year} {2019})},\ \Eprint {http://arxiv.org/abs/1810.10905}
  {arXiv:1810.10905 [hep-ph]} \BibitemShut {NoStop}%
\bibitem [{\citenamefont {Ellis}\ \emph {et~al.}(2020)\citenamefont {Ellis},
  \citenamefont {Evans}, \citenamefont {Nagata}, \citenamefont {Olive},\ and\
  \citenamefont {Velasco-Sevilla}}]{cmssm4}%
  \BibitemOpen
  \bibfield  {author} {\bibinfo {author} {\bibfnamefont {J.}~\bibnamefont
  {Ellis}}, \bibinfo {author} {\bibfnamefont {J.~L.}\ \bibnamefont {Evans}},
  \bibinfo {author} {\bibfnamefont {N.}~\bibnamefont {Nagata}}, \bibinfo
  {author} {\bibfnamefont {K.~A.}\ \bibnamefont {Olive}}, \ and\ \bibinfo
  {author} {\bibfnamefont {L.}~\bibnamefont {Velasco-Sevilla}},\ }\href
  {\doibase 10.1140/epjc/s10052-020-7872-3} {\bibfield  {journal} {\bibinfo
  {journal} {Eur. Phys. J. C}\ }\textbf {\bibinfo {volume} {80}},\ \bibinfo
  {pages} {332} (\bibinfo {year} {2020})},\ \Eprint
  {http://arxiv.org/abs/1912.04888} {arXiv:1912.04888 [hep-ph]} \BibitemShut
  {NoStop}%
\bibitem [{\citenamefont {Bagnaschi}\ \emph {et~al.}(2015)\citenamefont
  {Bagnaschi} \emph {et~al.}}]{mc12}%
  \BibitemOpen
  \bibfield  {author} {\bibinfo {author} {\bibfnamefont {E.~A.}\ \bibnamefont
  {Bagnaschi}} \emph {et~al.},\ }\href {\doibase
  10.1140/epjc/s10052-015-3718-9} {\bibfield  {journal} {\bibinfo  {journal}
  {Eur. Phys. J.}\ }\textbf {\bibinfo {volume} {C75}},\ \bibinfo {pages} {500}
  (\bibinfo {year} {2015})},\ \Eprint {http://arxiv.org/abs/1508.01173}
  {arXiv:1508.01173 [hep-ph]} \BibitemShut {NoStop}%
\bibitem [{\citenamefont {Bechtle}\ \emph {et~al.}(2016)\citenamefont {Bechtle}
  \emph {et~al.}}]{fittino}%
  \BibitemOpen
  \bibfield  {author} {\bibinfo {author} {\bibfnamefont {P.}~\bibnamefont
  {Bechtle}} \emph {et~al.},\ }\href {\doibase 10.1140/epjc/s10052-015-3864-0}
  {\bibfield  {journal} {\bibinfo  {journal} {Eur. Phys. J.}\ }\textbf
  {\bibinfo {volume} {C76}},\ \bibinfo {pages} {96} (\bibinfo {year} {2016})},\
  \Eprint {http://arxiv.org/abs/1508.05951} {arXiv:1508.05951 [hep-ph]}
  \BibitemShut {NoStop}%
\bibitem [{\citenamefont {Aaboud}\ \emph
  {et~al.}(2018{\natexlab{a}})\citenamefont {Aaboud} \emph {et~al.}}]{nosusy1}%
  \BibitemOpen
  \bibfield  {author} {\bibinfo {author} {\bibfnamefont {M.}~\bibnamefont
  {Aaboud}} \emph {et~al.} (\bibinfo {collaboration} {ATLAS}),\ }\href
  {\doibase 10.1007/JHEP06(2018)107} {\bibfield  {journal} {\bibinfo  {journal}
  {JHEP}\ }\textbf {\bibinfo {volume} {06}},\ \bibinfo {pages} {107} (\bibinfo
  {year} {2018}{\natexlab{a}})},\ \Eprint {http://arxiv.org/abs/1711.01901}
  {arXiv:1711.01901 [hep-ex]} \BibitemShut {NoStop}%
\bibitem [{\citenamefont {Aaboud}\ \emph
  {et~al.}(2018{\natexlab{b}})\citenamefont {Aaboud} \emph {et~al.}}]{nosusy2}%
  \BibitemOpen
  \bibfield  {author} {\bibinfo {author} {\bibfnamefont {M.}~\bibnamefont
  {Aaboud}} \emph {et~al.} (\bibinfo {collaboration} {ATLAS}),\ }\href
  {\doibase 10.1103/PhysRevD.97.112001} {\bibfield  {journal} {\bibinfo
  {journal} {Phys. Rev.}\ }\textbf {\bibinfo {volume} {D97}},\ \bibinfo {pages}
  {112001} (\bibinfo {year} {2018}{\natexlab{b}})},\ \Eprint
  {http://arxiv.org/abs/1712.02332} {arXiv:1712.02332 [hep-ex]} \BibitemShut
  {NoStop}%
\bibitem [{\citenamefont {Sirunyan}\ \emph {et~al.}(2017)\citenamefont
  {Sirunyan} \emph {et~al.}}]{nosusy3}%
  \BibitemOpen
  \bibfield  {author} {\bibinfo {author} {\bibfnamefont {A.~M.}\ \bibnamefont
  {Sirunyan}} \emph {et~al.} (\bibinfo {collaboration} {CMS}),\ }\href
  {\doibase 10.1140/epjc/s10052-017-5267-x} {\bibfield  {journal} {\bibinfo
  {journal} {Eur. Phys. J.}\ }\textbf {\bibinfo {volume} {C77}},\ \bibinfo
  {pages} {710} (\bibinfo {year} {2017})},\ \Eprint
  {http://arxiv.org/abs/1705.04650} {arXiv:1705.04650 [hep-ex]} \BibitemShut
  {NoStop}%
\bibitem [{\citenamefont {Sirunyan}\ \emph {et~al.}(2018)\citenamefont
  {Sirunyan} \emph {et~al.}}]{nosusy4}%
  \BibitemOpen
  \bibfield  {author} {\bibinfo {author} {\bibfnamefont {A.~M.}\ \bibnamefont
  {Sirunyan}} \emph {et~al.} (\bibinfo {collaboration} {CMS}),\ }\href
  {\doibase 10.1007/JHEP05(2018)025} {\bibfield  {journal} {\bibinfo  {journal}
  {JHEP}\ }\textbf {\bibinfo {volume} {05}},\ \bibinfo {pages} {025} (\bibinfo
  {year} {2018})},\ \Eprint {http://arxiv.org/abs/1802.02110} {arXiv:1802.02110
  [hep-ex]} \BibitemShut {NoStop}%
\bibitem [{\citenamefont {Hut}(1977)}]{hlw1}%
  \BibitemOpen
  \bibfield  {author} {\bibinfo {author} {\bibfnamefont {P.}~\bibnamefont
  {Hut}},\ }\href {\doibase 10.1016/0370-2693(77)90139-3} {\bibfield  {journal}
  {\bibinfo  {journal} {Phys. Lett.}\ }\textbf {\bibinfo {volume} {69B}},\
  \bibinfo {pages} {85} (\bibinfo {year} {1977})},\ \bibinfo {note}
  {[,179(1977)]}\BibitemShut {NoStop}%
\bibitem [{\citenamefont {Lee}\ and\ \citenamefont {Weinberg}(1977)}]{hlw2}%
  \BibitemOpen
  \bibfield  {author} {\bibinfo {author} {\bibfnamefont {B.~W.}\ \bibnamefont
  {Lee}}\ and\ \bibinfo {author} {\bibfnamefont {S.}~\bibnamefont {Weinberg}},\
  }\href {\doibase 10.1103/PhysRevLett.39.165} {\bibfield  {journal} {\bibinfo
  {journal} {Phys. Rev. Lett.}\ }\textbf {\bibinfo {volume} {39}},\ \bibinfo
  {pages} {165} (\bibinfo {year} {1977})},\ \bibinfo {note}
  {[,183(1977)]}\BibitemShut {NoStop}%
\bibitem [{\citenamefont {Hall}\ \emph {et~al.}(2010)\citenamefont {Hall},
  \citenamefont {Jedamzik}, \citenamefont {March-Russell},\ and\ \citenamefont
  {West}}]{fimp1}%
  \BibitemOpen
  \bibfield  {author} {\bibinfo {author} {\bibfnamefont {L.~J.}\ \bibnamefont
  {Hall}}, \bibinfo {author} {\bibfnamefont {K.}~\bibnamefont {Jedamzik}},
  \bibinfo {author} {\bibfnamefont {J.}~\bibnamefont {March-Russell}}, \ and\
  \bibinfo {author} {\bibfnamefont {S.~M.}\ \bibnamefont {West}},\ }\href
  {\doibase 10.1007/JHEP03(2010)080} {\bibfield  {journal} {\bibinfo  {journal}
  {JHEP}\ }\textbf {\bibinfo {volume} {03}},\ \bibinfo {pages} {080} (\bibinfo
  {year} {2010})},\ \Eprint {http://arxiv.org/abs/0911.1120} {arXiv:0911.1120
  [hep-ph]} \BibitemShut {NoStop}%
\bibitem [{\citenamefont {Bernal}\ \emph {et~al.}(2017)\citenamefont {Bernal},
  \citenamefont {Heikinheimo}, \citenamefont {Tenkanen}, \citenamefont
  {Tuominen},\ and\ \citenamefont {Vaskonen}}]{Bernal:2017kxu}%
  \BibitemOpen
  \bibfield  {author} {\bibinfo {author} {\bibfnamefont {N.}~\bibnamefont
  {Bernal}}, \bibinfo {author} {\bibfnamefont {M.}~\bibnamefont {Heikinheimo}},
  \bibinfo {author} {\bibfnamefont {T.}~\bibnamefont {Tenkanen}}, \bibinfo
  {author} {\bibfnamefont {K.}~\bibnamefont {Tuominen}}, \ and\ \bibinfo
  {author} {\bibfnamefont {V.}~\bibnamefont {Vaskonen}},\ }\href {\doibase
  10.1142/S0217751X1730023X} {\bibfield  {journal} {\bibinfo  {journal} {Int.
  J. Mod. Phys.}\ }\textbf {\bibinfo {volume} {A32}},\ \bibinfo {pages}
  {1730023} (\bibinfo {year} {2017})},\ \Eprint
  {http://arxiv.org/abs/1706.07442} {arXiv:1706.07442 [hep-ph]} \BibitemShut
  {NoStop}%
\bibitem [{\citenamefont {Pagels}\ and\ \citenamefont
  {Primack}(1982)}]{gravitino1}%
  \BibitemOpen
  \bibfield  {author} {\bibinfo {author} {\bibfnamefont {H.}~\bibnamefont
  {Pagels}}\ and\ \bibinfo {author} {\bibfnamefont {J.~R.}\ \bibnamefont
  {Primack}},\ }\href {\doibase 10.1103/PhysRevLett.48.223} {\bibfield
  {journal} {\bibinfo  {journal} {Phys. Rev. Lett.}\ }\textbf {\bibinfo
  {volume} {48}},\ \bibinfo {pages} {223} (\bibinfo {year} {1982})}\BibitemShut
  {NoStop}%
\bibitem [{\citenamefont {Nanopoulos}\ \emph {et~al.}(1983)\citenamefont
  {Nanopoulos}, \citenamefont {Olive},\ and\ \citenamefont
  {Srednicki}}]{gravitino2}%
  \BibitemOpen
  \bibfield  {author} {\bibinfo {author} {\bibfnamefont {D.~V.}\ \bibnamefont
  {Nanopoulos}}, \bibinfo {author} {\bibfnamefont {K.~A.}\ \bibnamefont
  {Olive}}, \ and\ \bibinfo {author} {\bibfnamefont {M.}~\bibnamefont
  {Srednicki}},\ }\href {\doibase 10.1016/0370-2693(83)91624-6} {\bibfield
  {journal} {\bibinfo  {journal} {Phys. Lett.}\ }\textbf {\bibinfo {volume}
  {127B}},\ \bibinfo {pages} {30} (\bibinfo {year} {1983})}\BibitemShut
  {NoStop}%
\bibitem [{\citenamefont {Khlopov}\ and\ \citenamefont
  {Linde}(1984)}]{gravitino3}%
  \BibitemOpen
  \bibfield  {author} {\bibinfo {author} {\bibfnamefont {M.~{\relax Yu}.}\
  \bibnamefont {Khlopov}}\ and\ \bibinfo {author} {\bibfnamefont {A.~D.}\
  \bibnamefont {Linde}},\ }\href {\doibase 10.1016/0370-2693(84)91656-3}
  {\bibfield  {journal} {\bibinfo  {journal} {Phys. Lett.}\ }\textbf {\bibinfo
  {volume} {138B}},\ \bibinfo {pages} {265} (\bibinfo {year}
  {1984})}\BibitemShut {NoStop}%
\bibitem [{\citenamefont {Feng}\ \emph {et~al.}(2004)\citenamefont {Feng},
  \citenamefont {Su},\ and\ \citenamefont {Takayama}}]{gravitino7}%
  \BibitemOpen
  \bibfield  {author} {\bibinfo {author} {\bibfnamefont {J.~L.}\ \bibnamefont
  {Feng}}, \bibinfo {author} {\bibfnamefont {S.}~\bibnamefont {Su}}, \ and\
  \bibinfo {author} {\bibfnamefont {F.}~\bibnamefont {Takayama}},\ }\href
  {\doibase 10.1103/PhysRevD.70.075019} {\bibfield  {journal} {\bibinfo
  {journal} {Phys. Rev.}\ }\textbf {\bibinfo {volume} {D70}},\ \bibinfo {pages}
  {075019} (\bibinfo {year} {2004})},\ \Eprint
  {http://arxiv.org/abs/hep-ph/0404231} {arXiv:hep-ph/0404231 [hep-ph]}
  \BibitemShut {NoStop}%
\bibitem [{\citenamefont {Steffen}(2006)}]{gravitino8}%
  \BibitemOpen
  \bibfield  {author} {\bibinfo {author} {\bibfnamefont {F.~D.}\ \bibnamefont
  {Steffen}},\ }\href {\doibase 10.1088/1475-7516/2006/09/001} {\bibfield
  {journal} {\bibinfo  {journal} {JCAP}\ }\textbf {\bibinfo {volume} {0609}},\
  \bibinfo {pages} {001} (\bibinfo {year} {2006})},\ \Eprint
  {http://arxiv.org/abs/hep-ph/0605306} {arXiv:hep-ph/0605306 [hep-ph]}
  \BibitemShut {NoStop}%
\bibitem [{\citenamefont {Covi}\ \emph {et~al.}(2009)\citenamefont {Covi},
  \citenamefont {Hasenkamp}, \citenamefont {Pokorski},\ and\ \citenamefont
  {Roberts}}]{gravitino12}%
  \BibitemOpen
  \bibfield  {author} {\bibinfo {author} {\bibfnamefont {L.}~\bibnamefont
  {Covi}}, \bibinfo {author} {\bibfnamefont {J.}~\bibnamefont {Hasenkamp}},
  \bibinfo {author} {\bibfnamefont {S.}~\bibnamefont {Pokorski}}, \ and\
  \bibinfo {author} {\bibfnamefont {J.}~\bibnamefont {Roberts}},\ }\href
  {\doibase 10.1088/1126-6708/2009/11/003} {\bibfield  {journal} {\bibinfo
  {journal} {JHEP}\ }\textbf {\bibinfo {volume} {11}},\ \bibinfo {pages} {003}
  (\bibinfo {year} {2009})},\ \Eprint {http://arxiv.org/abs/0908.3399}
  {arXiv:0908.3399 [hep-ph]} \BibitemShut {NoStop}%
\bibitem [{\citenamefont {Mambrini}\ \emph {et~al.}(2013)\citenamefont
  {Mambrini}, \citenamefont {Olive}, \citenamefont {Quevillon},\ and\
  \citenamefont {Zaldivar}}]{Mambrini:2013iaa}%
  \BibitemOpen
  \bibfield  {author} {\bibinfo {author} {\bibfnamefont {Y.}~\bibnamefont
  {Mambrini}}, \bibinfo {author} {\bibfnamefont {K.~A.}\ \bibnamefont {Olive}},
  \bibinfo {author} {\bibfnamefont {J.}~\bibnamefont {Quevillon}}, \ and\
  \bibinfo {author} {\bibfnamefont {B.}~\bibnamefont {Zaldivar}},\ }\href
  {\doibase 10.1103/PhysRevLett.110.241306} {\bibfield  {journal} {\bibinfo
  {journal} {Phys. Rev. Lett.}\ }\textbf {\bibinfo {volume} {110}},\ \bibinfo
  {pages} {241306} (\bibinfo {year} {2013})},\ \Eprint
  {http://arxiv.org/abs/1302.4438} {arXiv:1302.4438 [hep-ph]} \BibitemShut
  {NoStop}%
\bibitem [{\citenamefont {Mambrini}\ \emph {et~al.}(2015)\citenamefont
  {Mambrini}, \citenamefont {Nagata}, \citenamefont {Olive}, \citenamefont
  {Quevillon},\ and\ \citenamefont {Zheng}}]{mnoqz1}%
  \BibitemOpen
  \bibfield  {author} {\bibinfo {author} {\bibfnamefont {Y.}~\bibnamefont
  {Mambrini}}, \bibinfo {author} {\bibfnamefont {N.}~\bibnamefont {Nagata}},
  \bibinfo {author} {\bibfnamefont {K.~A.}\ \bibnamefont {Olive}}, \bibinfo
  {author} {\bibfnamefont {J.}~\bibnamefont {Quevillon}}, \ and\ \bibinfo
  {author} {\bibfnamefont {J.}~\bibnamefont {Zheng}},\ }\href {\doibase
  10.1103/PhysRevD.91.095010} {\bibfield  {journal} {\bibinfo  {journal} {Phys.
  Rev.}\ }\textbf {\bibinfo {volume} {D91}},\ \bibinfo {pages} {095010}
  (\bibinfo {year} {2015})},\ \Eprint {http://arxiv.org/abs/1502.06929}
  {arXiv:1502.06929 [hep-ph]} \BibitemShut {NoStop}%
\bibitem [{\citenamefont {Bhattacharyya}\ \emph {et~al.}(2018)\citenamefont
  {Bhattacharyya}, \citenamefont {Dutra}, \citenamefont {Mambrini},\ and\
  \citenamefont {Pierre}}]{Bhattacharyya:2018evo}%
  \BibitemOpen
  \bibfield  {author} {\bibinfo {author} {\bibfnamefont {G.}~\bibnamefont
  {Bhattacharyya}}, \bibinfo {author} {\bibfnamefont {M.}~\bibnamefont
  {Dutra}}, \bibinfo {author} {\bibfnamefont {Y.}~\bibnamefont {Mambrini}}, \
  and\ \bibinfo {author} {\bibfnamefont {M.}~\bibnamefont {Pierre}},\ }\href
  {\doibase 10.1103/PhysRevD.98.035038} {\bibfield  {journal} {\bibinfo
  {journal} {Phys. Rev.}\ }\textbf {\bibinfo {volume} {D98}},\ \bibinfo {pages}
  {035038} (\bibinfo {year} {2018})},\ \Eprint
  {http://arxiv.org/abs/1806.00016} {arXiv:1806.00016 [hep-ph]} \BibitemShut
  {NoStop}%
\bibitem [{\citenamefont {Benakli}\ \emph {et~al.}(2017)\citenamefont
  {Benakli}, \citenamefont {Chen}, \citenamefont {Dudas},\ and\ \citenamefont
  {Mambrini}}]{Benakli:2017whb}%
  \BibitemOpen
  \bibfield  {author} {\bibinfo {author} {\bibfnamefont {K.}~\bibnamefont
  {Benakli}}, \bibinfo {author} {\bibfnamefont {Y.}~\bibnamefont {Chen}},
  \bibinfo {author} {\bibfnamefont {E.}~\bibnamefont {Dudas}}, \ and\ \bibinfo
  {author} {\bibfnamefont {Y.}~\bibnamefont {Mambrini}},\ }\href {\doibase
  10.1103/PhysRevD.95.095002} {\bibfield  {journal} {\bibinfo  {journal} {Phys.
  Rev.}\ }\textbf {\bibinfo {volume} {D95}},\ \bibinfo {pages} {095002}
  (\bibinfo {year} {2017})},\ \Eprint {http://arxiv.org/abs/1701.06574}
  {arXiv:1701.06574 [hep-ph]} \BibitemShut {NoStop}%
\bibitem [{\citenamefont {Dudas}\ \emph
  {et~al.}(2017{\natexlab{a}})\citenamefont {Dudas}, \citenamefont {Mambrini},\
  and\ \citenamefont {Olive}}]{grav2}%
  \BibitemOpen
  \bibfield  {author} {\bibinfo {author} {\bibfnamefont {E.}~\bibnamefont
  {Dudas}}, \bibinfo {author} {\bibfnamefont {Y.}~\bibnamefont {Mambrini}}, \
  and\ \bibinfo {author} {\bibfnamefont {K.}~\bibnamefont {Olive}},\ }\href
  {\doibase 10.1103/PhysRevLett.119.051801} {\bibfield  {journal} {\bibinfo
  {journal} {Phys. Rev. Lett.}\ }\textbf {\bibinfo {volume} {119}},\ \bibinfo
  {pages} {051801} (\bibinfo {year} {2017}{\natexlab{a}})},\ \Eprint
  {http://arxiv.org/abs/1704.03008} {arXiv:1704.03008 [hep-ph]} \BibitemShut
  {NoStop}%
\bibitem [{\citenamefont {Dudas}\ \emph
  {et~al.}(2017{\natexlab{b}})\citenamefont {Dudas}, \citenamefont
  {Gherghetta}, \citenamefont {Mambrini},\ and\ \citenamefont {Olive}}]{grav3}%
  \BibitemOpen
  \bibfield  {author} {\bibinfo {author} {\bibfnamefont {E.}~\bibnamefont
  {Dudas}}, \bibinfo {author} {\bibfnamefont {T.}~\bibnamefont {Gherghetta}},
  \bibinfo {author} {\bibfnamefont {Y.}~\bibnamefont {Mambrini}}, \ and\
  \bibinfo {author} {\bibfnamefont {K.~A.}\ \bibnamefont {Olive}},\ }\href
  {\doibase 10.1103/PhysRevD.96.115032} {\bibfield  {journal} {\bibinfo
  {journal} {Phys. Rev.}\ }\textbf {\bibinfo {volume} {D96}},\ \bibinfo {pages}
  {115032} (\bibinfo {year} {2017}{\natexlab{b}})},\ \Eprint
  {http://arxiv.org/abs/1710.07341} {arXiv:1710.07341 [hep-ph]} \BibitemShut
  {NoStop}%
\bibitem [{\citenamefont {Ellis}\ \emph {et~al.}(2018)\citenamefont {Ellis},
  \citenamefont {Gherghetta}, \citenamefont {Kaneta},\ and\ \citenamefont
  {Olive}}]{highsc}%
  \BibitemOpen
  \bibfield  {author} {\bibinfo {author} {\bibfnamefont {S.~A.~R.}\
  \bibnamefont {Ellis}}, \bibinfo {author} {\bibfnamefont {T.}~\bibnamefont
  {Gherghetta}}, \bibinfo {author} {\bibfnamefont {K.}~\bibnamefont {Kaneta}},
  \ and\ \bibinfo {author} {\bibfnamefont {K.~A.}\ \bibnamefont {Olive}},\
  }\href {\doibase 10.1103/PhysRevD.98.055009} {\bibfield  {journal} {\bibinfo
  {journal} {Phys. Rev.}\ }\textbf {\bibinfo {volume} {D98}},\ \bibinfo {pages}
  {055009} (\bibinfo {year} {2018})},\ \Eprint
  {http://arxiv.org/abs/1807.06488} {arXiv:1807.06488 [hep-ph]} \BibitemShut
  {NoStop}%
\bibitem [{\citenamefont {Kaneta}\ \emph {et~al.}(2019)\citenamefont {Kaneta},
  \citenamefont {Mambrini},\ and\ \citenamefont {Olive}}]{Kaneta:2019zgw}%
  \BibitemOpen
  \bibfield  {author} {\bibinfo {author} {\bibfnamefont {K.}~\bibnamefont
  {Kaneta}}, \bibinfo {author} {\bibfnamefont {Y.}~\bibnamefont {Mambrini}}, \
  and\ \bibinfo {author} {\bibfnamefont {K.~A.}\ \bibnamefont {Olive}},\ }\href
  {\doibase 10.1103/PhysRevD.99.063508} {\bibfield  {journal} {\bibinfo
  {journal} {Phys. Rev.}\ }\textbf {\bibinfo {volume} {D99}},\ \bibinfo {pages}
  {063508} (\bibinfo {year} {2019})},\ \Eprint
  {http://arxiv.org/abs/1901.04449} {arXiv:1901.04449 [hep-ph]} \BibitemShut
  {NoStop}%
\bibitem [{\citenamefont {Bernal}\ \emph {et~al.}(2018)\citenamefont {Bernal},
  \citenamefont {Dutra}, \citenamefont {Mambrini}, \citenamefont {Olive},
  \citenamefont {Peloso},\ and\ \citenamefont {Pierre}}]{Bernal:2018qlk}%
  \BibitemOpen
  \bibfield  {author} {\bibinfo {author} {\bibfnamefont {N.}~\bibnamefont
  {Bernal}}, \bibinfo {author} {\bibfnamefont {M.}~\bibnamefont {Dutra}},
  \bibinfo {author} {\bibfnamefont {Y.}~\bibnamefont {Mambrini}}, \bibinfo
  {author} {\bibfnamefont {K.}~\bibnamefont {Olive}}, \bibinfo {author}
  {\bibfnamefont {M.}~\bibnamefont {Peloso}}, \ and\ \bibinfo {author}
  {\bibfnamefont {M.}~\bibnamefont {Pierre}},\ }\href {\doibase
  10.1103/PhysRevD.97.115020} {\bibfield  {journal} {\bibinfo  {journal} {Phys.
  Rev.}\ }\textbf {\bibinfo {volume} {D97}},\ \bibinfo {pages} {115020}
  (\bibinfo {year} {2018})},\ \Eprint {http://arxiv.org/abs/1803.01866}
  {arXiv:1803.01866 [hep-ph]} \BibitemShut {NoStop}%
\bibitem [{\citenamefont {Chowdhury}\ \emph {et~al.}(2019)\citenamefont
  {Chowdhury}, \citenamefont {Dudas}, \citenamefont {Dutra},\ and\
  \citenamefont {Mambrini}}]{Chowdhury:2018tzw}%
  \BibitemOpen
  \bibfield  {author} {\bibinfo {author} {\bibfnamefont {D.}~\bibnamefont
  {Chowdhury}}, \bibinfo {author} {\bibfnamefont {E.}~\bibnamefont {Dudas}},
  \bibinfo {author} {\bibfnamefont {M.}~\bibnamefont {Dutra}}, \ and\ \bibinfo
  {author} {\bibfnamefont {Y.}~\bibnamefont {Mambrini}},\ }\href {\doibase
  10.1103/PhysRevD.99.095028} {\bibfield  {journal} {\bibinfo  {journal} {Phys.
  Rev.}\ }\textbf {\bibinfo {volume} {D99}},\ \bibinfo {pages} {095028}
  (\bibinfo {year} {2019})},\ \Eprint {http://arxiv.org/abs/1811.01947}
  {arXiv:1811.01947 [hep-ph]} \BibitemShut {NoStop}%
\bibitem [{\citenamefont {Giudice}\ \emph {et~al.}(2001)\citenamefont
  {Giudice}, \citenamefont {Kolb},\ and\ \citenamefont
  {Riotto}}]{Giudice:2000ex}%
  \BibitemOpen
  \bibfield  {author} {\bibinfo {author} {\bibfnamefont {G.~F.}\ \bibnamefont
  {Giudice}}, \bibinfo {author} {\bibfnamefont {E.~W.}\ \bibnamefont {Kolb}}, \
  and\ \bibinfo {author} {\bibfnamefont {A.}~\bibnamefont {Riotto}},\ }\href
  {\doibase 10.1103/PhysRevD.64.023508} {\bibfield  {journal} {\bibinfo
  {journal} {Phys. Rev.}\ }\textbf {\bibinfo {volume} {D64}},\ \bibinfo {pages}
  {023508} (\bibinfo {year} {2001})},\ \Eprint
  {http://arxiv.org/abs/hep-ph/0005123} {arXiv:hep-ph/0005123 [hep-ph]}
  \BibitemShut {NoStop}%
\bibitem [{\citenamefont {Chung}\ \emph {et~al.}(1999)\citenamefont {Chung},
  \citenamefont {Kolb},\ and\ \citenamefont {Riotto}}]{Chung:1998rq}%
  \BibitemOpen
  \bibfield  {author} {\bibinfo {author} {\bibfnamefont {D.~J.~H.}\
  \bibnamefont {Chung}}, \bibinfo {author} {\bibfnamefont {E.~W.}\ \bibnamefont
  {Kolb}}, \ and\ \bibinfo {author} {\bibfnamefont {A.}~\bibnamefont
  {Riotto}},\ }\href {\doibase 10.1103/PhysRevD.60.063504} {\bibfield
  {journal} {\bibinfo  {journal} {Phys. Rev.}\ }\textbf {\bibinfo {volume}
  {D60}},\ \bibinfo {pages} {063504} (\bibinfo {year} {1999})},\ \Eprint
  {http://arxiv.org/abs/hep-ph/9809453} {arXiv:hep-ph/9809453 [hep-ph]}
  \BibitemShut {NoStop}%
\bibitem [{\citenamefont {Garcia}\ \emph {et~al.}(2017)\citenamefont {Garcia},
  \citenamefont {Mambrini}, \citenamefont {Olive},\ and\ \citenamefont
  {Peloso}}]{Garcia:2017tuj}%
  \BibitemOpen
  \bibfield  {author} {\bibinfo {author} {\bibfnamefont {M.~A.~G.}\
  \bibnamefont {Garcia}}, \bibinfo {author} {\bibfnamefont {Y.}~\bibnamefont
  {Mambrini}}, \bibinfo {author} {\bibfnamefont {K.~A.}\ \bibnamefont {Olive}},
  \ and\ \bibinfo {author} {\bibfnamefont {M.}~\bibnamefont {Peloso}},\ }\href
  {\doibase 10.1103/PhysRevD.96.103510} {\bibfield  {journal} {\bibinfo
  {journal} {Phys. Rev.}\ }\textbf {\bibinfo {volume} {D96}},\ \bibinfo {pages}
  {103510} (\bibinfo {year} {2017})},\ \Eprint
  {http://arxiv.org/abs/1709.01549} {arXiv:1709.01549 [hep-ph]} \BibitemShut
  {NoStop}%
\bibitem [{\citenamefont {Chen}\ and\ \citenamefont
  {Kang}(2018)}]{Chen:2017kvz}%
  \BibitemOpen
  \bibfield  {author} {\bibinfo {author} {\bibfnamefont {S.-L.}\ \bibnamefont
  {Chen}}\ and\ \bibinfo {author} {\bibfnamefont {Z.}~\bibnamefont {Kang}},\
  }\href {\doibase 10.1088/1475-7516/2018/05/036} {\bibfield  {journal}
  {\bibinfo  {journal} {JCAP}\ }\textbf {\bibinfo {volume} {1805}},\ \bibinfo
  {pages} {036} (\bibinfo {year} {2018})},\ \Eprint
  {http://arxiv.org/abs/1711.02556} {arXiv:1711.02556 [hep-ph]} \BibitemShut
  {NoStop}%
\bibitem [{\citenamefont {Harigaya}\ and\ \citenamefont
  {Mukaida}(2014)}]{Harigaya:2013vwa}%
  \BibitemOpen
  \bibfield  {author} {\bibinfo {author} {\bibfnamefont {K.}~\bibnamefont
  {Harigaya}}\ and\ \bibinfo {author} {\bibfnamefont {K.}~\bibnamefont
  {Mukaida}},\ }\href {\doibase 10.1007/JHEP05(2014)006} {\bibfield  {journal}
  {\bibinfo  {journal} {JHEP}\ }\textbf {\bibinfo {volume} {05}},\ \bibinfo
  {pages} {006} (\bibinfo {year} {2014})},\ \Eprint
  {http://arxiv.org/abs/1312.3097} {arXiv:1312.3097 [hep-ph]} \BibitemShut
  {NoStop}%
\bibitem [{\citenamefont {Harigaya}\ \emph {et~al.}(2014)\citenamefont
  {Harigaya}, \citenamefont {Kawasaki}, \citenamefont {Mukaida},\ and\
  \citenamefont {Yamada}}]{Harigaya:2014waa}%
  \BibitemOpen
  \bibfield  {author} {\bibinfo {author} {\bibfnamefont {K.}~\bibnamefont
  {Harigaya}}, \bibinfo {author} {\bibfnamefont {M.}~\bibnamefont {Kawasaki}},
  \bibinfo {author} {\bibfnamefont {K.}~\bibnamefont {Mukaida}}, \ and\
  \bibinfo {author} {\bibfnamefont {M.}~\bibnamefont {Yamada}},\ }\href
  {\doibase 10.1103/PhysRevD.89.083532} {\bibfield  {journal} {\bibinfo
  {journal} {Phys. Rev. D}\ }\textbf {\bibinfo {volume} {89}},\ \bibinfo
  {pages} {083532} (\bibinfo {year} {2014})},\ \Eprint
  {http://arxiv.org/abs/1402.2846} {arXiv:1402.2846 [hep-ph]} \BibitemShut
  {NoStop}%
\bibitem [{\citenamefont {Mukaida}\ and\ \citenamefont
  {Yamada}(2016)}]{Mukaida:2015ria}%
  \BibitemOpen
  \bibfield  {author} {\bibinfo {author} {\bibfnamefont {K.}~\bibnamefont
  {Mukaida}}\ and\ \bibinfo {author} {\bibfnamefont {M.}~\bibnamefont
  {Yamada}},\ }\href {\doibase 10.1088/1475-7516/2016/02/003} {\bibfield
  {journal} {\bibinfo  {journal} {JCAP}\ }\textbf {\bibinfo {volume} {1602}},\
  \bibinfo {pages} {003} (\bibinfo {year} {2016})},\ \Eprint
  {http://arxiv.org/abs/1506.07661} {arXiv:1506.07661 [hep-ph]} \BibitemShut
  {NoStop}%
\bibitem [{\citenamefont {Garcia}\ and\ \citenamefont
  {Amin}(2018)}]{Garcia:2018wtq}%
  \BibitemOpen
  \bibfield  {author} {\bibinfo {author} {\bibfnamefont {M.~A.~G.}\
  \bibnamefont {Garcia}}\ and\ \bibinfo {author} {\bibfnamefont {M.~A.}\
  \bibnamefont {Amin}},\ }\href {\doibase 10.1103/PhysRevD.98.103504}
  {\bibfield  {journal} {\bibinfo  {journal} {Phys. Rev.}\ }\textbf {\bibinfo
  {volume} {D98}},\ \bibinfo {pages} {103504} (\bibinfo {year} {2018})},\
  \Eprint {http://arxiv.org/abs/1806.01865} {arXiv:1806.01865 [hep-ph]}
  \BibitemShut {NoStop}%
\bibitem [{\citenamefont {Harigaya}\ \emph {et~al.}(2019)\citenamefont
  {Harigaya}, \citenamefont {Mukaida},\ and\ \citenamefont
  {Yamada}}]{Harigaya:2019tzu}%
  \BibitemOpen
  \bibfield  {author} {\bibinfo {author} {\bibfnamefont {K.}~\bibnamefont
  {Harigaya}}, \bibinfo {author} {\bibfnamefont {K.}~\bibnamefont {Mukaida}}, \
  and\ \bibinfo {author} {\bibfnamefont {M.}~\bibnamefont {Yamada}},\ }\href
  {\doibase 10.1007/JHEP07(2019)059} {\bibfield  {journal} {\bibinfo  {journal}
  {JHEP}\ }\textbf {\bibinfo {volume} {07}},\ \bibinfo {pages} {059} (\bibinfo
  {year} {2019})},\ \Eprint {http://arxiv.org/abs/1901.11027} {arXiv:1901.11027
  [hep-ph]} \BibitemShut {NoStop}%
\bibitem [{\citenamefont {Bernal}\ \emph {et~al.}(2019)\citenamefont {Bernal},
  \citenamefont {Elahi}, \citenamefont {Maldonado},\ and\ \citenamefont
  {Unwin}}]{Bernal:2019mhf}%
  \BibitemOpen
  \bibfield  {author} {\bibinfo {author} {\bibfnamefont {N.}~\bibnamefont
  {Bernal}}, \bibinfo {author} {\bibfnamefont {F.}~\bibnamefont {Elahi}},
  \bibinfo {author} {\bibfnamefont {C.}~\bibnamefont {Maldonado}}, \ and\
  \bibinfo {author} {\bibfnamefont {J.}~\bibnamefont {Unwin}},\ }\href
  {\doibase 10.1088/1475-7516/2019/11/026} {\bibfield  {journal} {\bibinfo
  {journal} {JCAP}\ }\textbf {\bibinfo {volume} {1911}},\ \bibinfo {pages}
  {026} (\bibinfo {year} {2019})},\ \Eprint {http://arxiv.org/abs/1909.07992}
  {arXiv:1909.07992 [hep-ph]} \BibitemShut {NoStop}%
\bibitem [{\citenamefont {Kallosh}\ and\ \citenamefont
  {Linde}(2013{\natexlab{a}})}]{Kallosh:2013hoa}%
  \BibitemOpen
  \bibfield  {author} {\bibinfo {author} {\bibfnamefont {R.}~\bibnamefont
  {Kallosh}}\ and\ \bibinfo {author} {\bibfnamefont {A.}~\bibnamefont
  {Linde}},\ }\href {\doibase 10.1088/1475-7516/2013/07/002} {\bibfield
  {journal} {\bibinfo  {journal} {JCAP}\ }\textbf {\bibinfo {volume} {1307}},\
  \bibinfo {pages} {002} (\bibinfo {year} {2013}{\natexlab{a}})},\ \Eprint
  {http://arxiv.org/abs/1306.5220} {arXiv:1306.5220 [hep-th]} \BibitemShut
  {NoStop}%
\bibitem [{\citenamefont {Starobinsky}(1980)}]{Staro}%
  \BibitemOpen
  \bibfield  {author} {\bibinfo {author} {\bibfnamefont {A.~A.}\ \bibnamefont
  {Starobinsky}},\ }\href {\doibase 10.1016/0370-2693(80)90670-X} {\bibfield
  {journal} {\bibinfo  {journal} {Phys. Lett.}\ }\textbf {\bibinfo {volume}
  {91B}},\ \bibinfo {pages} {99} (\bibinfo {year} {1980})},\ \bibinfo {note}
  {[Adv. Ser. Astrophys. Cosmol.3,130(1987); ,771(1980)]}\BibitemShut {NoStop}%
\bibitem [{\citenamefont {Cremmer}\ \emph {et~al.}(1983)\citenamefont
  {Cremmer}, \citenamefont {Ferrara}, \citenamefont {Kounnas},\ and\
  \citenamefont {Nanopoulos}}]{no-scale1}%
  \BibitemOpen
  \bibfield  {author} {\bibinfo {author} {\bibfnamefont {E.}~\bibnamefont
  {Cremmer}}, \bibinfo {author} {\bibfnamefont {S.}~\bibnamefont {Ferrara}},
  \bibinfo {author} {\bibfnamefont {C.}~\bibnamefont {Kounnas}}, \ and\
  \bibinfo {author} {\bibfnamefont {D.~V.}\ \bibnamefont {Nanopoulos}},\ }\href
  {\doibase 10.1016/0370-2693(83)90106-5} {\bibfield  {journal} {\bibinfo
  {journal} {Phys. Lett.}\ }\textbf {\bibinfo {volume} {133B}},\ \bibinfo
  {pages} {61} (\bibinfo {year} {1983})}\BibitemShut {NoStop}%
\bibitem [{\citenamefont {Ellis}\ \emph
  {et~al.}(1984{\natexlab{b}})\citenamefont {Ellis}, \citenamefont {Lahanas},
  \citenamefont {Nanopoulos},\ and\ \citenamefont {Tamvakis}}]{no-scale2}%
  \BibitemOpen
  \bibfield  {author} {\bibinfo {author} {\bibfnamefont {J.~R.}\ \bibnamefont
  {Ellis}}, \bibinfo {author} {\bibfnamefont {A.~B.}\ \bibnamefont {Lahanas}},
  \bibinfo {author} {\bibfnamefont {D.~V.}\ \bibnamefont {Nanopoulos}}, \ and\
  \bibinfo {author} {\bibfnamefont {K.}~\bibnamefont {Tamvakis}},\ }\href
  {\doibase 10.1016/0370-2693(84)91378-9} {\bibfield  {journal} {\bibinfo
  {journal} {Phys. Lett.}\ }\textbf {\bibinfo {volume} {134B}},\ \bibinfo
  {pages} {429} (\bibinfo {year} {1984}{\natexlab{b}})}\BibitemShut {NoStop}%
\bibitem [{\citenamefont {Lahanas}\ and\ \citenamefont
  {Nanopoulos}(1987)}]{LN}%
  \BibitemOpen
  \bibfield  {author} {\bibinfo {author} {\bibfnamefont {A.}~\bibnamefont
  {Lahanas}}\ and\ \bibinfo {author} {\bibfnamefont {D.~V.}\ \bibnamefont
  {Nanopoulos}},\ }\href {\doibase 10.1016/0370-1573(87)90034-2} {\bibfield
  {journal} {\bibinfo  {journal} {Phys.\ Rept.}\ }\textbf {\bibinfo {volume}
  {145}},\ \bibinfo {pages} {1} (\bibinfo {year} {1987})}\BibitemShut {NoStop}%
\bibitem [{\citenamefont {Ellis}\ \emph {et~al.}(2015)\citenamefont {Ellis},
  \citenamefont {Garc\'ia}, \citenamefont {Nanopoulos},\ and\ \citenamefont
  {Olive}}]{Ellis:2015pla}%
  \BibitemOpen
  \bibfield  {author} {\bibinfo {author} {\bibfnamefont {J.}~\bibnamefont
  {Ellis}}, \bibinfo {author} {\bibfnamefont {M.~A.~G.}\ \bibnamefont
  {Garc\'ia}}, \bibinfo {author} {\bibfnamefont {D.~V.}\ \bibnamefont
  {Nanopoulos}}, \ and\ \bibinfo {author} {\bibfnamefont {K.~A.}\ \bibnamefont
  {Olive}},\ }\href {\doibase 10.1088/1475-7516/2015/07/050} {\bibfield
  {journal} {\bibinfo  {journal} {JCAP}\ }\textbf {\bibinfo {volume} {1507}},\
  \bibinfo {pages} {050} (\bibinfo {year} {2015})},\ \Eprint
  {http://arxiv.org/abs/1505.06986} {arXiv:1505.06986 [hep-ph]} \BibitemShut
  {NoStop}%
\bibitem [{\citenamefont {Shtanov}\ \emph {et~al.}(1995)\citenamefont
  {Shtanov}, \citenamefont {Traschen},\ and\ \citenamefont
  {Brandenberger}}]{Shtanov:1994ce}%
  \BibitemOpen
  \bibfield  {author} {\bibinfo {author} {\bibfnamefont {Y.}~\bibnamefont
  {Shtanov}}, \bibinfo {author} {\bibfnamefont {J.~H.}\ \bibnamefont
  {Traschen}}, \ and\ \bibinfo {author} {\bibfnamefont {R.~H.}\ \bibnamefont
  {Brandenberger}},\ }\href {\doibase 10.1103/PhysRevD.51.5438} {\bibfield
  {journal} {\bibinfo  {journal} {Phys. Rev.}\ }\textbf {\bibinfo {volume}
  {D51}},\ \bibinfo {pages} {5438} (\bibinfo {year} {1995})},\ \Eprint
  {http://arxiv.org/abs/hep-ph/9407247} {arXiv:hep-ph/9407247 [hep-ph]}
  \BibitemShut {NoStop}%
\bibitem [{\citenamefont {Ichikawa}\ \emph {et~al.}(2008)\citenamefont
  {Ichikawa}, \citenamefont {Suyama}, \citenamefont {Takahashi},\ and\
  \citenamefont {Yamaguchi}}]{Ichikawa:2008ne}%
  \BibitemOpen
  \bibfield  {author} {\bibinfo {author} {\bibfnamefont {K.}~\bibnamefont
  {Ichikawa}}, \bibinfo {author} {\bibfnamefont {T.}~\bibnamefont {Suyama}},
  \bibinfo {author} {\bibfnamefont {T.}~\bibnamefont {Takahashi}}, \ and\
  \bibinfo {author} {\bibfnamefont {M.}~\bibnamefont {Yamaguchi}},\ }\href
  {\doibase 10.1103/PhysRevD.78.063545} {\bibfield  {journal} {\bibinfo
  {journal} {Phys. Rev.}\ }\textbf {\bibinfo {volume} {D78}},\ \bibinfo {pages}
  {063545} (\bibinfo {year} {2008})},\ \Eprint {http://arxiv.org/abs/0807.3988}
  {arXiv:0807.3988 [astro-ph]} \BibitemShut {NoStop}%
\bibitem [{\citenamefont {Chu}\ \emph {et~al.}(2014)\citenamefont {Chu},
  \citenamefont {Mambrini}, \citenamefont {Quevillon},\ and\ \citenamefont
  {Zaldivar}}]{fimp3}%
  \BibitemOpen
  \bibfield  {author} {\bibinfo {author} {\bibfnamefont {X.}~\bibnamefont
  {Chu}}, \bibinfo {author} {\bibfnamefont {Y.}~\bibnamefont {Mambrini}},
  \bibinfo {author} {\bibfnamefont {J.}~\bibnamefont {Quevillon}}, \ and\
  \bibinfo {author} {\bibfnamefont {B.}~\bibnamefont {Zaldivar}},\ }\href
  {\doibase 10.1088/1475-7516/2014/01/034} {\bibfield  {journal} {\bibinfo
  {journal} {JCAP}\ }\textbf {\bibinfo {volume} {1401}},\ \bibinfo {pages}
  {034} (\bibinfo {year} {2014})},\ \Eprint {http://arxiv.org/abs/1306.4677}
  {arXiv:1306.4677 [hep-ph]} \BibitemShut {NoStop}%
\bibitem [{\citenamefont {Nagata}\ \emph {et~al.}(2015)\citenamefont {Nagata},
  \citenamefont {Olive},\ and\ \citenamefont {Zheng}}]{Nagata:2015dma}%
  \BibitemOpen
  \bibfield  {author} {\bibinfo {author} {\bibfnamefont {N.}~\bibnamefont
  {Nagata}}, \bibinfo {author} {\bibfnamefont {K.~A.}\ \bibnamefont {Olive}}, \
  and\ \bibinfo {author} {\bibfnamefont {J.}~\bibnamefont {Zheng}},\ }\href
  {\doibase 10.1007/JHEP10(2015)193} {\bibfield  {journal} {\bibinfo  {journal}
  {JHEP}\ }\textbf {\bibinfo {volume} {10}},\ \bibinfo {pages} {193} (\bibinfo
  {year} {2015})},\ \Eprint {http://arxiv.org/abs/1509.00809} {arXiv:1509.00809
  [hep-ph]} \BibitemShut {NoStop}%
\bibitem [{\citenamefont {Mambrini}\ \emph {et~al.}(2016)\citenamefont
  {Mambrini}, \citenamefont {Nagata}, \citenamefont {Olive},\ and\
  \citenamefont {Zheng}}]{Mambrini:2016dca}%
  \BibitemOpen
  \bibfield  {author} {\bibinfo {author} {\bibfnamefont {Y.}~\bibnamefont
  {Mambrini}}, \bibinfo {author} {\bibfnamefont {N.}~\bibnamefont {Nagata}},
  \bibinfo {author} {\bibfnamefont {K.~A.}\ \bibnamefont {Olive}}, \ and\
  \bibinfo {author} {\bibfnamefont {J.}~\bibnamefont {Zheng}},\ }\href
  {\doibase 10.1103/PhysRevD.93.111703} {\bibfield  {journal} {\bibinfo
  {journal} {Phys. Rev.}\ }\textbf {\bibinfo {volume} {D93}},\ \bibinfo {pages}
  {111703} (\bibinfo {year} {2016})},\ \Eprint
  {http://arxiv.org/abs/1602.05583} {arXiv:1602.05583 [hep-ph]} \BibitemShut
  {NoStop}%
\bibitem [{\citenamefont {Nagata}\ \emph {et~al.}(2017)\citenamefont {Nagata},
  \citenamefont {Olive},\ and\ \citenamefont {Zheng}}]{mnoqz2}%
  \BibitemOpen
  \bibfield  {author} {\bibinfo {author} {\bibfnamefont {N.}~\bibnamefont
  {Nagata}}, \bibinfo {author} {\bibfnamefont {K.~A.}\ \bibnamefont {Olive}}, \
  and\ \bibinfo {author} {\bibfnamefont {J.}~\bibnamefont {Zheng}},\ }\href
  {\doibase 10.1088/1475-7516/2017/02/016} {\bibfield  {journal} {\bibinfo
  {journal} {JCAP}\ }\textbf {\bibinfo {volume} {1702}},\ \bibinfo {pages}
  {016} (\bibinfo {year} {2017})},\ \Eprint {http://arxiv.org/abs/1611.04693}
  {arXiv:1611.04693 [hep-ph]} \BibitemShut {NoStop}%
\bibitem [{\citenamefont {Tanabashi}\ \emph {et~al.}(2018)\citenamefont
  {Tanabashi} \emph {et~al.}}]{Tanabashi:2018oca}%
  \BibitemOpen
  \bibfield  {author} {\bibinfo {author} {\bibfnamefont {M.}~\bibnamefont
  {Tanabashi}} \emph {et~al.} (\bibinfo {collaboration} {Particle Data
  Group}),\ }\href {\doibase 10.1103/PhysRevD.98.030001} {\bibfield  {journal}
  {\bibinfo  {journal} {Phys. Rev.}\ }\textbf {\bibinfo {volume} {D98}},\
  \bibinfo {pages} {030001} (\bibinfo {year} {2018})}\BibitemShut {NoStop}%
\bibitem [{\citenamefont {Roszkowski}\ \emph {et~al.}(2014)\citenamefont
  {Roszkowski}, \citenamefont {Trojanowski},\ and\ \citenamefont
  {Turzy\'nski}}]{Roszkowski:2014lga}%
  \BibitemOpen
  \bibfield  {author} {\bibinfo {author} {\bibfnamefont {L.}~\bibnamefont
  {Roszkowski}}, \bibinfo {author} {\bibfnamefont {S.}~\bibnamefont
  {Trojanowski}}, \ and\ \bibinfo {author} {\bibfnamefont {K.}~\bibnamefont
  {Turzy\'nski}},\ }\href {\doibase 10.1007/JHEP11(2014)146} {\bibfield
  {journal} {\bibinfo  {journal} {JHEP}\ }\textbf {\bibinfo {volume} {11}},\
  \bibinfo {pages} {146} (\bibinfo {year} {2014})},\ \Eprint
  {http://arxiv.org/abs/1406.0012} {arXiv:1406.0012 [hep-ph]} \BibitemShut
  {NoStop}%
\bibitem [{\citenamefont {Dudas}\ \emph {et~al.}(2018)\citenamefont {Dudas},
  \citenamefont {Gherghetta}, \citenamefont {Kaneta}, \citenamefont
  {Mambrini},\ and\ \citenamefont {Olive}}]{grav4}%
  \BibitemOpen
  \bibfield  {author} {\bibinfo {author} {\bibfnamefont {E.}~\bibnamefont
  {Dudas}}, \bibinfo {author} {\bibfnamefont {T.}~\bibnamefont {Gherghetta}},
  \bibinfo {author} {\bibfnamefont {K.}~\bibnamefont {Kaneta}}, \bibinfo
  {author} {\bibfnamefont {Y.}~\bibnamefont {Mambrini}}, \ and\ \bibinfo
  {author} {\bibfnamefont {K.~A.}\ \bibnamefont {Olive}},\ }\href {\doibase
  10.1103/PhysRevD.98.015030} {\bibfield  {journal} {\bibinfo  {journal} {Phys.
  Rev.}\ }\textbf {\bibinfo {volume} {D98}},\ \bibinfo {pages} {015030}
  (\bibinfo {year} {2018})},\ \Eprint {http://arxiv.org/abs/1805.07342}
  {arXiv:1805.07342 [hep-ph]} \BibitemShut {NoStop}%
\bibitem [{\citenamefont {Lozanov}\ and\ \citenamefont
  {Amin}(2017)}]{Lozanov:2016hid}%
  \BibitemOpen
  \bibfield  {author} {\bibinfo {author} {\bibfnamefont {K.~D.}\ \bibnamefont
  {Lozanov}}\ and\ \bibinfo {author} {\bibfnamefont {M.~A.}\ \bibnamefont
  {Amin}},\ }\href {\doibase 10.1103/PhysRevLett.119.061301} {\bibfield
  {journal} {\bibinfo  {journal} {Phys. Rev. Lett.}\ }\textbf {\bibinfo
  {volume} {119}},\ \bibinfo {pages} {061301} (\bibinfo {year} {2017})},\
  \Eprint {http://arxiv.org/abs/1608.01213} {arXiv:1608.01213 [astro-ph.CO]}
  \BibitemShut {NoStop}%
\bibitem [{\citenamefont {Kofman}\ \emph {et~al.}(1997)\citenamefont {Kofman},
  \citenamefont {Linde},\ and\ \citenamefont {Starobinsky}}]{Kofman:1997yn}%
  \BibitemOpen
  \bibfield  {author} {\bibinfo {author} {\bibfnamefont {L.}~\bibnamefont
  {Kofman}}, \bibinfo {author} {\bibfnamefont {A.}~\bibnamefont {Linde}}, \
  and\ \bibinfo {author} {\bibfnamefont {A.}~\bibnamefont {Starobinsky}},\
  }\href {\doibase 10.1103/PhysRevD.56.3258} {\bibfield  {journal} {\bibinfo
  {journal} {Phys. Rev.}\ }\textbf {\bibinfo {volume} {D56}},\ \bibinfo {pages}
  {3258} (\bibinfo {year} {1997})},\ \Eprint
  {http://arxiv.org/abs/hep-ph/9704452} {arXiv:hep-ph/9704452 [hep-ph]}
  \BibitemShut {NoStop}%
\bibitem [{\citenamefont {Greene}\ \emph {et~al.}(1997)\citenamefont {Greene},
  \citenamefont {Kofman}, \citenamefont {Linde},\ and\ \citenamefont
  {Starobinsky}}]{Greene:1997fu}%
  \BibitemOpen
  \bibfield  {author} {\bibinfo {author} {\bibfnamefont {P.~B.}\ \bibnamefont
  {Greene}}, \bibinfo {author} {\bibfnamefont {L.}~\bibnamefont {Kofman}},
  \bibinfo {author} {\bibfnamefont {A.~D.}\ \bibnamefont {Linde}}, \ and\
  \bibinfo {author} {\bibfnamefont {A.~A.}\ \bibnamefont {Starobinsky}},\
  }\href {\doibase 10.1103/PhysRevD.56.6175} {\bibfield  {journal} {\bibinfo
  {journal} {Phys. Rev.}\ }\textbf {\bibinfo {volume} {D56}},\ \bibinfo {pages}
  {6175} (\bibinfo {year} {1997})},\ \Eprint
  {http://arxiv.org/abs/hep-ph/9705347} {arXiv:hep-ph/9705347 [hep-ph]}
  \BibitemShut {NoStop}%
\bibitem [{\citenamefont {Felder}\ \emph {et~al.}(1999)\citenamefont {Felder},
  \citenamefont {Kofman},\ and\ \citenamefont {Linde}}]{Felder:1998vq}%
  \BibitemOpen
  \bibfield  {author} {\bibinfo {author} {\bibfnamefont {G.~N.}\ \bibnamefont
  {Felder}}, \bibinfo {author} {\bibfnamefont {L.}~\bibnamefont {Kofman}}, \
  and\ \bibinfo {author} {\bibfnamefont {A.~D.}\ \bibnamefont {Linde}},\ }\href
  {\doibase 10.1103/PhysRevD.59.123523} {\bibfield  {journal} {\bibinfo
  {journal} {Phys. Rev.}\ }\textbf {\bibinfo {volume} {D59}},\ \bibinfo {pages}
  {123523} (\bibinfo {year} {1999})},\ \Eprint
  {http://arxiv.org/abs/hep-ph/9812289} {arXiv:hep-ph/9812289 [hep-ph]}
  \BibitemShut {NoStop}%
\bibitem [{\citenamefont {Amin}\ \emph {et~al.}(2015)\citenamefont {Amin},
  \citenamefont {Hertzberg}, \citenamefont {Kaiser},\ and\ \citenamefont
  {Karouby}}]{Amin:2014eta}%
  \BibitemOpen
  \bibfield  {author} {\bibinfo {author} {\bibfnamefont {M.~A.}\ \bibnamefont
  {Amin}}, \bibinfo {author} {\bibfnamefont {M.~P.}\ \bibnamefont {Hertzberg}},
  \bibinfo {author} {\bibfnamefont {D.~I.}\ \bibnamefont {Kaiser}}, \ and\
  \bibinfo {author} {\bibfnamefont {J.}~\bibnamefont {Karouby}},\ }\href
  {\doibase 10.1142/S0218271815300037} {\bibfield  {journal} {\bibinfo
  {journal} {Int. J. Mod. Phys. D}\ }\textbf {\bibinfo {volume} {24}},\
  \bibinfo {pages} {1530003} (\bibinfo {year} {2015})},\ \Eprint
  {http://arxiv.org/abs/1410.3808} {arXiv:1410.3808 [hep-ph]} \BibitemShut
  {NoStop}%
\bibitem [{\citenamefont {Lozanov}\ and\ \citenamefont
  {Amin}(2018)}]{Lozanov:2017hjm}%
  \BibitemOpen
  \bibfield  {author} {\bibinfo {author} {\bibfnamefont {K.~D.}\ \bibnamefont
  {Lozanov}}\ and\ \bibinfo {author} {\bibfnamefont {M.~A.}\ \bibnamefont
  {Amin}},\ }\href {\doibase 10.1103/PhysRevD.97.023533} {\bibfield  {journal}
  {\bibinfo  {journal} {Phys. Rev.}\ }\textbf {\bibinfo {volume} {D97}},\
  \bibinfo {pages} {023533} (\bibinfo {year} {2018})},\ \Eprint
  {http://arxiv.org/abs/1710.06851} {arXiv:1710.06851 [astro-ph.CO]}
  \BibitemShut {NoStop}%
\bibitem [{\citenamefont {Akrami}\ \emph {et~al.}(2018)\citenamefont {Akrami}
  \emph {et~al.}}]{planck3}%
  \BibitemOpen
  \bibfield  {author} {\bibinfo {author} {\bibfnamefont {Y.}~\bibnamefont
  {Akrami}} \emph {et~al.} (\bibinfo {collaboration} {Planck}),\ }\href@noop {}
  {\  (\bibinfo {year} {2018})},\ \Eprint {http://arxiv.org/abs/1807.06211}
  {arXiv:1807.06211 [astro-ph.CO]} \BibitemShut {NoStop}%
\bibitem [{\citenamefont {Ade}\ \emph {et~al.}(2018)\citenamefont {Ade} \emph
  {et~al.}}]{Ade:2018gkx}%
  \BibitemOpen
  \bibfield  {author} {\bibinfo {author} {\bibfnamefont {P.~A.~R.}\
  \bibnamefont {Ade}} \emph {et~al.} (\bibinfo {collaboration} {BICEP2, Keck
  Array}),\ }\href {\doibase 10.1103/PhysRevLett.121.221301} {\bibfield
  {journal} {\bibinfo  {journal} {Phys. Rev. Lett.}\ }\textbf {\bibinfo
  {volume} {121}},\ \bibinfo {pages} {221301} (\bibinfo {year} {2018})},\
  \Eprint {http://arxiv.org/abs/1810.05216} {arXiv:1810.05216 [astro-ph.CO]}
  \BibitemShut {NoStop}%
\bibitem [{\citenamefont {Bezrukov}\ and\ \citenamefont
  {Shaposhnikov}(2008)}]{Bezrukov:2007ep}%
  \BibitemOpen
  \bibfield  {author} {\bibinfo {author} {\bibfnamefont {F.~L.}\ \bibnamefont
  {Bezrukov}}\ and\ \bibinfo {author} {\bibfnamefont {M.}~\bibnamefont
  {Shaposhnikov}},\ }\href {\doibase 10.1016/j.physletb.2007.11.072} {\bibfield
   {journal} {\bibinfo  {journal} {Phys. Lett.}\ }\textbf {\bibinfo {volume}
  {B659}},\ \bibinfo {pages} {703} (\bibinfo {year} {2008})},\ \Eprint
  {http://arxiv.org/abs/0710.3755} {arXiv:0710.3755 [hep-th]} \BibitemShut
  {NoStop}%
\bibitem [{\citenamefont {Ellis}\ \emph
  {et~al.}(2013{\natexlab{a}})\citenamefont {Ellis}, \citenamefont
  {Nanopoulos},\ and\ \citenamefont {Olive}}]{Ellis:2013nxa}%
  \BibitemOpen
  \bibfield  {author} {\bibinfo {author} {\bibfnamefont {J.}~\bibnamefont
  {Ellis}}, \bibinfo {author} {\bibfnamefont {D.~V.}\ \bibnamefont
  {Nanopoulos}}, \ and\ \bibinfo {author} {\bibfnamefont {K.~A.}\ \bibnamefont
  {Olive}},\ }\href {\doibase 10.1088/1475-7516/2013/10/009} {\bibfield
  {journal} {\bibinfo  {journal} {JCAP}\ }\textbf {\bibinfo {volume} {1310}},\
  \bibinfo {pages} {009} (\bibinfo {year} {2013}{\natexlab{a}})},\ \Eprint
  {http://arxiv.org/abs/1307.3537} {arXiv:1307.3537 [hep-th]} \BibitemShut
  {NoStop}%
\bibitem [{\citenamefont {Kallosh}\ and\ \citenamefont
  {Linde}(2013{\natexlab{b}})}]{Kallosh:2013daa}%
  \BibitemOpen
  \bibfield  {author} {\bibinfo {author} {\bibfnamefont {R.}~\bibnamefont
  {Kallosh}}\ and\ \bibinfo {author} {\bibfnamefont {A.}~\bibnamefont
  {Linde}},\ }\href {\doibase 10.1088/1475-7516/2013/12/006} {\bibfield
  {journal} {\bibinfo  {journal} {JCAP}\ }\textbf {\bibinfo {volume} {1312}},\
  \bibinfo {pages} {006} (\bibinfo {year} {2013}{\natexlab{b}})},\ \Eprint
  {http://arxiv.org/abs/1309.2015} {arXiv:1309.2015 [hep-th]} \BibitemShut
  {NoStop}%
\bibitem [{\citenamefont {Kallosh}\ \emph {et~al.}(2013)\citenamefont
  {Kallosh}, \citenamefont {Linde},\ and\ \citenamefont
  {Roest}}]{Kallosh:2013yoa}%
  \BibitemOpen
  \bibfield  {author} {\bibinfo {author} {\bibfnamefont {R.}~\bibnamefont
  {Kallosh}}, \bibinfo {author} {\bibfnamefont {A.}~\bibnamefont {Linde}}, \
  and\ \bibinfo {author} {\bibfnamefont {D.}~\bibnamefont {Roest}},\ }\href
  {\doibase 10.1007/JHEP11(2013)198} {\bibfield  {journal} {\bibinfo  {journal}
  {JHEP}\ }\textbf {\bibinfo {volume} {11}},\ \bibinfo {pages} {198} (\bibinfo
  {year} {2013})},\ \Eprint {http://arxiv.org/abs/1311.0472} {arXiv:1311.0472
  [hep-th]} \BibitemShut {NoStop}%
\bibitem [{\citenamefont {Ferrara}\ \emph {et~al.}(2013)\citenamefont
  {Ferrara}, \citenamefont {Kallosh}, \citenamefont {Linde},\ and\
  \citenamefont {Porrati}}]{Ferrara:2013rsa}%
  \BibitemOpen
  \bibfield  {author} {\bibinfo {author} {\bibfnamefont {S.}~\bibnamefont
  {Ferrara}}, \bibinfo {author} {\bibfnamefont {R.}~\bibnamefont {Kallosh}},
  \bibinfo {author} {\bibfnamefont {A.}~\bibnamefont {Linde}}, \ and\ \bibinfo
  {author} {\bibfnamefont {M.}~\bibnamefont {Porrati}},\ }\href {\doibase
  10.1103/PhysRevD.88.085038} {\bibfield  {journal} {\bibinfo  {journal} {Phys.
  Rev.}\ }\textbf {\bibinfo {volume} {D88}},\ \bibinfo {pages} {085038}
  (\bibinfo {year} {2013})},\ \Eprint {http://arxiv.org/abs/1307.7696}
  {arXiv:1307.7696 [hep-th]} \BibitemShut {NoStop}%
\bibitem [{\citenamefont {Kallosh}\ and\ \citenamefont
  {Linde}(2013{\natexlab{c}})}]{Kallosh:2013maa}%
  \BibitemOpen
  \bibfield  {author} {\bibinfo {author} {\bibfnamefont {R.}~\bibnamefont
  {Kallosh}}\ and\ \bibinfo {author} {\bibfnamefont {A.}~\bibnamefont
  {Linde}},\ }\href {\doibase 10.1088/1475-7516/2013/10/033} {\bibfield
  {journal} {\bibinfo  {journal} {JCAP}\ }\textbf {\bibinfo {volume} {1310}},\
  \bibinfo {pages} {033} (\bibinfo {year} {2013}{\natexlab{c}})},\ \Eprint
  {http://arxiv.org/abs/1307.7938} {arXiv:1307.7938 [hep-th]} \BibitemShut
  {NoStop}%
\bibitem [{\citenamefont {Kallosh}\ and\ \citenamefont
  {Linde}(2013{\natexlab{d}})}]{Kallosh:2013xya}%
  \BibitemOpen
  \bibfield  {author} {\bibinfo {author} {\bibfnamefont {R.}~\bibnamefont
  {Kallosh}}\ and\ \bibinfo {author} {\bibfnamefont {A.}~\bibnamefont
  {Linde}},\ }\href {\doibase 10.1088/1475-7516/2013/06/028} {\bibfield
  {journal} {\bibinfo  {journal} {JCAP}\ }\textbf {\bibinfo {volume} {1306}},\
  \bibinfo {pages} {028} (\bibinfo {year} {2013}{\natexlab{d}})},\ \Eprint
  {http://arxiv.org/abs/1306.3214} {arXiv:1306.3214 [hep-th]} \BibitemShut
  {NoStop}%
\bibitem [{\citenamefont {Ellis}\ \emph
  {et~al.}(2013{\natexlab{b}})\citenamefont {Ellis}, \citenamefont
  {Nanopoulos},\ and\ \citenamefont {Olive}}]{ENO6}%
  \BibitemOpen
  \bibfield  {author} {\bibinfo {author} {\bibfnamefont {J.}~\bibnamefont
  {Ellis}}, \bibinfo {author} {\bibfnamefont {D.~V.}\ \bibnamefont
  {Nanopoulos}}, \ and\ \bibinfo {author} {\bibfnamefont {K.~A.}\ \bibnamefont
  {Olive}},\ }\href {\doibase 10.1103/PhysRevLett.111.129902,
  10.1103/PhysRevLett.111.111301} {\bibfield  {journal} {\bibinfo  {journal}
  {Phys. Rev. Lett.}\ }\textbf {\bibinfo {volume} {111}},\ \bibinfo {pages}
  {111301} (\bibinfo {year} {2013}{\natexlab{b}})},\ \bibinfo {note} {[Erratum:
  Phys. Rev. Lett.111,no.12,129902(2013)]},\ \Eprint
  {http://arxiv.org/abs/1305.1247} {arXiv:1305.1247 [hep-th]} \BibitemShut
  {NoStop}%
\bibitem [{\citenamefont {Ellis}\ \emph {et~al.}(2019)\citenamefont {Ellis},
  \citenamefont {Nanopoulos}, \citenamefont {Olive},\ and\ \citenamefont
  {Verner}}]{Ellis:2018zya}%
  \BibitemOpen
  \bibfield  {author} {\bibinfo {author} {\bibfnamefont {J.}~\bibnamefont
  {Ellis}}, \bibinfo {author} {\bibfnamefont {D.~V.}\ \bibnamefont
  {Nanopoulos}}, \bibinfo {author} {\bibfnamefont {K.~A.}\ \bibnamefont
  {Olive}}, \ and\ \bibinfo {author} {\bibfnamefont {S.}~\bibnamefont
  {Verner}},\ }\href {\doibase 10.1007/JHEP03(2019)099} {\bibfield  {journal}
  {\bibinfo  {journal} {JHEP}\ }\textbf {\bibinfo {volume} {03}},\ \bibinfo
  {pages} {099} (\bibinfo {year} {2019})},\ \Eprint
  {http://arxiv.org/abs/1812.02192} {arXiv:1812.02192 [hep-th]} \BibitemShut
  {NoStop}%
\bibitem [{\citenamefont {Cecotti}(1987)}]{Cecotti}%
  \BibitemOpen
  \bibfield  {author} {\bibinfo {author} {\bibfnamefont {S.}~\bibnamefont
  {Cecotti}},\ }\href {\doibase 10.1016/0370-2693(87)90844-6} {\bibfield
  {journal} {\bibinfo  {journal} {Phys. Lett.}\ }\textbf {\bibinfo {volume}
  {B190}},\ \bibinfo {pages} {86} (\bibinfo {year} {1987})}\BibitemShut
  {NoStop}%
\bibitem [{\citenamefont {Aghanim}\ \emph {et~al.}(2018)\citenamefont {Aghanim}
  \emph {et~al.}}]{planck2}%
  \BibitemOpen
  \bibfield  {author} {\bibinfo {author} {\bibfnamefont {N.}~\bibnamefont
  {Aghanim}} \emph {et~al.} (\bibinfo {collaboration} {Planck}),\ }\href@noop
  {} {\  (\bibinfo {year} {2018})},\ \Eprint {http://arxiv.org/abs/1807.06209}
  {arXiv:1807.06209 [astro-ph.CO]} \BibitemShut {NoStop}%
\bibitem [{\citenamefont {Martin}\ and\ \citenamefont
  {Ringeval}(2010)}]{Martin:2010kz}%
  \BibitemOpen
  \bibfield  {author} {\bibinfo {author} {\bibfnamefont {J.}~\bibnamefont
  {Martin}}\ and\ \bibinfo {author} {\bibfnamefont {C.}~\bibnamefont
  {Ringeval}},\ }\href {\doibase 10.1103/PhysRevD.82.023511} {\bibfield
  {journal} {\bibinfo  {journal} {Phys. Rev.}\ }\textbf {\bibinfo {volume}
  {D82}},\ \bibinfo {pages} {023511} (\bibinfo {year} {2010})},\ \Eprint
  {http://arxiv.org/abs/1004.5525} {arXiv:1004.5525 [astro-ph.CO]} \BibitemShut
  {NoStop}%
\bibitem [{\citenamefont {Liddle}\ and\ \citenamefont
  {Leach}(2003)}]{Liddle:2003as}%
  \BibitemOpen
  \bibfield  {author} {\bibinfo {author} {\bibfnamefont {A.~R.}\ \bibnamefont
  {Liddle}}\ and\ \bibinfo {author} {\bibfnamefont {S.~M.}\ \bibnamefont
  {Leach}},\ }\href {\doibase 10.1103/PhysRevD.68.103503} {\bibfield  {journal}
  {\bibinfo  {journal} {Phys. Rev.}\ }\textbf {\bibinfo {volume} {D68}},\
  \bibinfo {pages} {103503} (\bibinfo {year} {2003})},\ \Eprint
  {http://arxiv.org/abs/astro-ph/0305263} {arXiv:astro-ph/0305263 [astro-ph]}
  \BibitemShut {NoStop}%
\bibitem [{\citenamefont {Fixsen}(2009)}]{Fixsen:2009ug}%
  \BibitemOpen
  \bibfield  {author} {\bibinfo {author} {\bibfnamefont {D.~J.}\ \bibnamefont
  {Fixsen}},\ }\href {\doibase 10.1088/0004-637X/707/2/916} {\bibfield
  {journal} {\bibinfo  {journal} {Astrophys. J.}\ }\textbf {\bibinfo {volume}
  {707}},\ \bibinfo {pages} {916} (\bibinfo {year} {2009})},\ \Eprint
  {http://arxiv.org/abs/0911.1955} {arXiv:0911.1955 [astro-ph.CO]} \BibitemShut
  {NoStop}%
\bibitem [{\citenamefont {Hasegawa}\ \emph {et~al.}(2019)\citenamefont
  {Hasegawa}, \citenamefont {Hiroshima}, \citenamefont {Kohri}, \citenamefont
  {Hansen}, \citenamefont {Tram},\ and\ \citenamefont
  {Hannestad}}]{Hasegawa:2019jsa}%
  \BibitemOpen
  \bibfield  {author} {\bibinfo {author} {\bibfnamefont {T.}~\bibnamefont
  {Hasegawa}}, \bibinfo {author} {\bibfnamefont {N.}~\bibnamefont {Hiroshima}},
  \bibinfo {author} {\bibfnamefont {K.}~\bibnamefont {Kohri}}, \bibinfo
  {author} {\bibfnamefont {R.~S.}\ \bibnamefont {Hansen}}, \bibinfo {author}
  {\bibfnamefont {T.}~\bibnamefont {Tram}}, \ and\ \bibinfo {author}
  {\bibfnamefont {S.}~\bibnamefont {Hannestad}},\ }\href {\doibase
  10.1088/1475-7516/2019/12/012} {\bibfield  {journal} {\bibinfo  {journal}
  {JCAP}\ }\textbf {\bibinfo {volume} {12}},\ \bibinfo {pages} {012} (\bibinfo
  {year} {2019})},\ \Eprint {http://arxiv.org/abs/1908.10189} {arXiv:1908.10189
  [hep-ph]} \BibitemShut {NoStop}%
\bibitem [{\citenamefont {Fields}\ \emph {et~al.}(2020)\citenamefont {Fields},
  \citenamefont {Olive}, \citenamefont {Yeh},\ and\ \citenamefont
  {Young}}]{Fields:2019pfx}%
  \BibitemOpen
  \bibfield  {author} {\bibinfo {author} {\bibfnamefont {B.~D.}\ \bibnamefont
  {Fields}}, \bibinfo {author} {\bibfnamefont {K.~A.}\ \bibnamefont {Olive}},
  \bibinfo {author} {\bibfnamefont {T.-H.}\ \bibnamefont {Yeh}}, \ and\
  \bibinfo {author} {\bibfnamefont {C.}~\bibnamefont {Young}},\ }\href
  {\doibase 10.1088/1475-7516/2020/03/010} {\bibfield  {journal} {\bibinfo
  {journal} {JCAP}\ }\textbf {\bibinfo {volume} {03}},\ \bibinfo {pages} {010}
  (\bibinfo {year} {2020})},\ \Eprint {http://arxiv.org/abs/1912.01132}
  {arXiv:1912.01132 [astro-ph.CO]} \BibitemShut {NoStop}%
\end{thebibliography}%

\end{document}